\pdfoutput=1
\documentclass[letterpaper]{JHEP3}[11pt]

\usepackage{amsmath}
\usepackage{amssymb}
\usepackage{amsfonts}
\usepackage{graphicx}
\usepackage{caption}
\usepackage{subcaption}
\usepackage{amsmath}
\usepackage{amssymb}
\usepackage{amsfonts}
\usepackage{lscape, graphicx}
\usepackage{cite}       
\usepackage{bm}         
\usepackage{verbatim}
\usepackage{subfig}

\usepackage{amsmath}
\usepackage{amssymb}
\usepackage{amsfonts}
\usepackage{lscape, graphicx}
\usepackage{cite}       
  
\let\ifpdf\relax

\usepackage[all]{xy}
\advance\parskip 1.0pt plus 1.0pt minus 2.0pt   
\def\href#1#2{#2}	

\def\N{{\cal N}}

\def\R{{\mathbb R}}
\def\S{{\mathbb S}}

\def\tr{{\rm tr}}

\def\Z{{\mathbb Z}}

\def\Dslash{{\rlap{\raise 1pt \hbox{$\>/$}}D}}

\def\lsim{\mathrel {\vcenter {\baselineskip 0pt \kern 0pt
    \hbox{$<$} \kern 0pt \hbox{$\sim$} }}}
\def\gsim{\mathrel {\vcenter {\baselineskip 0pt \kern 0pt
    \hbox{$>$} \kern 0pt \hbox{$\sim$} }}}

\def\cB{{\mathcal B}}
\def\cM{{\mathcal M}}

\def\gf{\mathfrak{g}}

\def\ad{\text{ad}}
\def\Gad{{G_\ad}}

\def\bc{\begin{center}}
\def\ec{\end{center}}
 
\title{Universal mechanism of (semi-classical) deconfinement
and $\mathbf{\theta}$-dependence for all simple groups}

\author   
    {
    {
    \def\href#1#2{#2}	
    Erich Poppitz,$^{1,}$\footnote{\email{poppitz@physics.utoronto.ca}} ~
    Thomas Sch\" afer,$^{2,}$\footnote{\email{tmschaef@ncsu.edu}}  ~
    Mithat \"Unsal$^{\; 3,}$\footnote{\email{unsal@sfsu.edu}}~
   \\${}^1$Department of Physics, University of Toronto,   
     Toronto, ON M5S 1A7, Canada
   \\${}^2$Department of Physics, North Carolina State University, 
     Raleigh, NC 27695, USA
   \\${}^3$Department of Physics and Astronomy, SFSU, 
     San Francisco, CA 94132, USA\\
        }
    }%

\abstract{
Using the twisted partition function on $\R^3 \times \S^1$, we argue 
that the deconfinement phase transition in pure Yang-Mills theory 
for all simple gauge groups is continuously connected to a quantum 
phase transition that can be studied in a controlled way. We explicitly 
consider two classes of theories, gauge theories with a center symmetry,  
such as $SU(N_c)$ gauge theory for arbitrary $N_c$, and theories without 
a center symmetry, such as $G_2$ gauge theory. The mechanism governing 
the phase transition is  universal and valid for all simple groups.
The perturbative one-loop potential as well as monopole-instantons 
generate attraction among the eigenvalues of the Wilson line. This is 
counter-acted by neutral bions --- topological excitations which generate  
eigenvalue repulsion for all simple groups. The transition is driven by 
the competition between these three effects. We study the transition 
in more detail for the gauge groups $SU(N_c)$, $N_c \geq 3$, and $G_2$. 
In the case of $G_2$ there is no change of symmetry, but the expectation 
value of the Wilson line exhibits a discontinuity.  We also examine the 
effect of the $\theta$-angle on the phase transition and critical temperature 
$T_c(\theta)$. The critical temperature is a multi-branched function,
which has a minimum at $\theta=\pi$ as a result of topological 
interference.}

\smallskip

\begin{document}

\section{Introduction}

 In our recent work, we studied a semi-classical mechanism for the 
center symmetry changing phase transition in four dimensional  
non-abelian $SU(2)$ gauge theory \cite{Poppitz:2012sw}.  The main idea 
of that work was to use continuity and the notion of a twisted partition 
function. We argued that the deconfining phase transition of pure 
Yang-Mills theory on $\R^3 \times \S^1_\beta$, which takes place at 
strong coupling, is continuously connected to a quantum phase transition 
that can be studied reliably, albeit using non-perturbative methods, at 
weak coupling. 

 The most important non-perturbative effect that arises in the weak 
coupling realization of the center symmetry changing phase transition is 
governed by a new class of topological objects, called  ``neutral bions". 
These objects are related to four dimensional instantons, but their 
physical effects are quite different. The 't Hooft vertex for a neutral 
bion induces a center-stabilizing potential  in $SU(2)$ theory, by generating  
a repulsion among the eigenvalues of the Wilson line.\footnote{For this 
reason, we sometimes refer to the neutral bion in theories with a 
center symmetry as ``center-stabilizing bion" \cite{Poppitz:2011wy}.} 
This effect counter-acts the center-destabilizing perturbative potential  
for the Wilson line, which leads to an attraction among the eigenvalues 
\cite{Gross:1980br}. In our previous work we showed that the 
monopole-instanton amplitudes also lead to an attraction among the 
eigenvalues of Wilson line. We systematically studied the competition 
between these three effects in the case of $SU(2)$ gauge theory. We 
observed that the twisted partition function on $\R^3 \times \S^1$ 
exhibits a second order phase transition, consistent with the second 
order transition observed in lattice gauge simulations of 
thermal $SU(2)$ Yang Mills theory.

\begin{table}[t]
\begin{center}
\begin{tabular}{|c|c|c|c|} \hline
$\gf$   & \phantom{\framebox{$G$}}$G$\phantom{\framebox{$G$}}  
  &  $\Gad :=G/Z(G)$ &$\pi_1(\Gad)=Z(G)$ \\ \hline\hline
$A_{N-1}$ 	& $\text{\it SU}(N)$    	& $\text{\it PSU}(N)$
  & $\Z_N$\\
$B_N$	& $\text{\it Spin}(2N+1)$& $\text{\it SO}(2N+1)$    
  & $\Z_2$\\
$C_N$	&  $\text{\it Sp}(N)$ or $\text{\it USp}(2N)$ & $\text{\it PSp}(N)$ 
  & $\Z_2$\\
$D_{2N}$	& $\text{\it Spin}(4N)$	& $\text{\it PSO}(4N)$    
  & $\Z_2\times\Z_2$\\
$D_{2N+1}$& $\text{\it Spin}(4N+2)$& $\text{\it PSO}(4N+2)$
  & $\Z_4$\\
$E_6$	& $E_6$    	& $E_6^{-78}$  & $\Z_3$   \\ 
$E_7$	& $E_7$	        & $E_7^{-133}$ & $\Z_2$  \\ 
$E_8$	& $E_8$    	& $E_8$       & $1$   \\ 
$F_4$ 	& $F_4$    	& $F_4$       & $1$  \\ 
$G_2$ 	& $G_2$   	& $G_2$       &  $1$  \\ 
\hline
\end{tabular}
\caption{\label{tab:lie}
The simple Lie algebras $\gf$ together with  their associated 
compact simply-connected Lie groups $G$ and the compact adjoint Lie 
groups $\Gad$. The last column lists the center symmetry of $G$, 
which is isomorphic to the fundamental group of  $\Gad$.}
\end{center}
\end{table}

\noindent In the present work, our goal is two-fold:
\begin{itemize} 
\item[{\it a)}] to generalize the phase transition mechanism in $SU(2)$ 
to  arbitrary compact simply-connected Lie  groups, $G$, including 
the theories without a center, $G_2, F_4, E_8$.  
\item[{\it b)}] to understand the topological $\theta$-angle dependence  
of the  critical temperature $T_c$. 
\end{itemize}

\noindent For convenience, we list the simple Lie algebras $\gf$ together 
with  their associated compact simply-connected Lie groups $G$ and the 
compact adjoint Lie groups $\Gad$ in Table \ref{tab:lie}. The last column 
lists the center group  of $G$, which is isomorphic to the fundamental 
group of  $\Gad$. For theories with adjoint matter, the center symmetry is 
the same as the center group.

\subsection{Twisted partition function and continuity of the deconfinement 
transition between weak and strong coupling}
\label{sec:twisted}

\begin{FIGURE}[ht]{
    \parbox[c]{\textwidth}
        {
        \begin{center}
        \includegraphics[angle=0, scale=0.5]{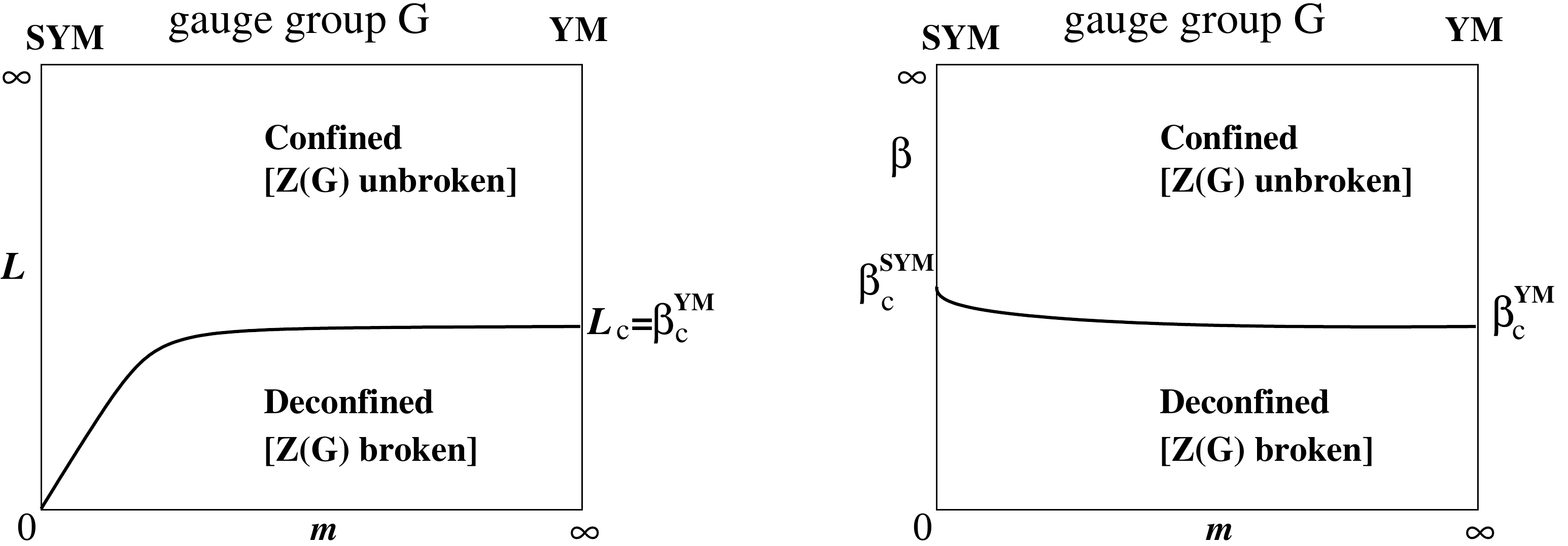}
	\hfil
        \caption
	{\small Phase diagram of Yang Mills (YM) theory with an 
        arbitrary simply-connected gauge group $G$ and a single
        Weyl fermion of mass $m$ in the adjoint representation,
        subject to periodic (left) and anti-periodic (right) boundary 
         conditions on $\R^3 \times \S^1$. 
         {\bf Right:} For the ordinary partition function the transition 
         occurs at a critical temperature of order $\Lambda$, and is not 
         accessible by analytical means. 
         {\bf Left:} In the case of the twisted partition function, the 
        thermal deconfinement phase transition in pure Yang Mills theory, 
        corresponding to the limit $m\to\infty$, is conjectured to be 
        continuously connected to a quantum phase transition in 
        supersymmetric Yang Mills (SYM) theory deformed by a gluino mass term. 
        For small $m$, the transition
        takes place as small $L$ and is analytically tractable. For 
        theories without center symmetry,  there is no symmetry 
        breaking associated with the transition. } 
        \label{fig:phasediag}
        \end{center}
        }
    }
\end{FIGURE}

 We  consider  Yang-Mills (YM) theory with one adjoint Weyl fermion 
with mass $m$.  In the chiral limit this theory reduces to  $\N=1$ 
supersymmetric YM theory,  and in the limit $m\to\infty$ it corresponds
to  pure YM theory.  Let $H$ and $F$ denote the Hamiltonian and  
fermion number operator, respectively, and $L$ the circumference of the 
$\S^1$ circle.  We introduce the twisted (non-thermal) partition 
function:
\begin{equation}
\label{tpf}
\widetilde Z ( L, m) = \tr \left[ e^{-L H} (-1)^F \right] 
  = \tr \left[ e^{-L H+ i \pi F}  \right]~,
\end{equation}
which, when viewed  as a  function of fermion mass, interpolates between 
the supersymmetric (Witten) index and the ordinary thermal partition 
function of the pure Yang-Mills theory. Namely, 
 \begin{align}
\label{limits}
&\widetilde Z ( L, m=0) =  I_{S} (G),   \cr 
 &\widetilde Z ( L, m=\infty) =  Z^{\rm YM} (\beta), \;\;  L\equiv \beta~,
\end{align}
where $I_{S} (G)=h^\vee $ is the index, which is equal to an invariant 
$h^\vee$ (independent of $L$) associated with the group  $G$ and called 
the dual Coxeter number ($h^\vee$ equals $N_c$ for $SU(N_c)$). Since 
$d \widetilde Z(L,0)/dL =0 $ for any $L$, there is no phase transition 
at $m=0$ \cite{Seiberg:1996nz, Aharony:1997bx,Davies:2000nw}. 
 $Z^{\rm YM} (\beta)$ is the  partition function of the pure 
Yang-Mills theory. Eqs.~(\ref{tpf})  and (\ref{limits}) permit  us to 
continue the deconfinement phase transition in pure YM theory to a 
semi-classically calculable transition in the small $L$-$m$ regime 
as shown in Fig.~\ref{fig:phasediag}.

\subsection{Review of the dynamics for $\mathbf{m=0}$ and small-$\mathbf{L}$}
\label{sec:review}

The dynamics of ${\cal  N}=1$ SYM theory with gauge group $G$ in the 
small-$L$ weak coupling  regime is that of the abelianized theory, 
$U(1)^r$, where $\; r=\text{rank}(\gf)$ is the rank of the Lie algebra 
of $G$  \cite{Seiberg:1996nz, Aharony:1997bx,Davies:2000nw}. The abelian 
description is valid at distance scales $\gtrsim  L r/2\pi$ 
\cite{Unsal:2010qh}. This follows from the fact that the Wilson line 
holonomy behaves as a compact  ``adjoint Higgs field" in the weak coupling 
regime. We note that there is no such description at strong coupling, 
see  the detailed discussion in Section~\ref{sec:weakstrong}. In a 
Euclidean context, the long distance theory is described  as a dilute 
gas of topological 1-defects (item (i) below) and 2-defects (items (ii) 
and (iii)):
\begin{itemize}
\item[{(i)}] monopole-instantons  $\cM_i$ with $i=1,\ldots, 
  \text{rank}(\gf)+1$,
\item[(ii)] magnetic bions  $\cB_{ij} = [\cM_i\overline \cM_j]$, 
 $\forall \;   \hat A_{ij} < 0$,    
\item[ (iii)] neutral bions   $\cB_{ii} = [\cM_i \overline \cM_i]$,  
 $\forall \;   \hat A_{ii} >0$. 
\end{itemize}
Here,  $\hat A_{ij}$ is the extended Cartan matrix of $G$.  Small 4d 
instantons are responsible for breaking the anomalous axial $U(1)_A$  
symmetry down to an anomaly free discrete subgroup, $\Z_{2 h^\vee}$. But 
apart from that, they have no sizable impact to any physical phenomena 
in this regime.  In the small-$L$ regime, 4d instantons can be viewed as 
composites of the monopole-instantons, i.e., ${\cal I}_{4d} \sim \prod_{j=0}^{r} 
[\cM_j]^{k_j^\vee}$, where $ k_j^\vee$ are dual Kac-labels (or co-marks; equal 
to unity for $SU(N_c)$)  and $\sum_{j=0}^r k_j^\vee = h^\vee= I_S(G)$ is the 
dual Coxeter number. This formula embodies the physics of fractionalization 
of a large 4d instanton $\gtrsim  Lr$ into  $k_j^\vee$ monopole-instantons 
of type $\cM_j$, each of which carry two adjoint fermionic 
zero-modes.\footnote{The topological classification of finite action
solutions on $\R^3\times\S^1$ can be found in \cite{Gross:1980br}. Monopole 
constituents of the instanton (caloron) solution for non-trivial holonomy 
were found in \cite{Lee:1997vp,Kraan:1998sn}, and the relevant index theorem 
is discussed in \cite{Nye:2000eg,Poppitz:2008hr}. The monopole vertex for 
general gauge group can be found in \cite{Davies:2000nw}. Magnetic bions 
are studied in \cite{Unsal:2007jx,Anber:2011de}, and neutral bions are 
discussed in \cite{Poppitz:2011wy,Poppitz:2012sw,Argyres:2012ka}.}

Since large 4d-instantons do not exist in this regime the problem of 
the breakdown of the dilute instanton gas approximation, which follows 
from the fact that instanton size is a  modulus, is avoided. Moreover,  
since the 1-defects and 2-defects obey a separation of scales,
\begin{align}  
r_\text{m} \ll r_\text{b} \ll d_\text{m-m} \ll  d_\text{b-b},
\end{align}
the treatment of monopole-instantons ($\cM_i$) and topological molecules 
($\cB_{ij}$) in the dilute-gas approximation and the effective theory derived 
from their proliferation in the vacuum is reliable. This is depicted in 
Fig.~\ref{fig:plasma}:  $r_\text{m}$ and  $r_\text{b}$ are the typical sizes  
of a monopole-instanton and bion events, and $d_\text{m-m}$ and $d_\text{b-b}$ 
are the typical  inter-monopole-instanton and  inter-bion separation.
For a detailed discussion of the scales, see \cite{Argyres:2012ka}. 
 
\begin{FIGURE}[ht]{
    \parbox[c]{\textwidth}
        {
        \begin{center}
        \includegraphics[angle=0, scale=0.5]{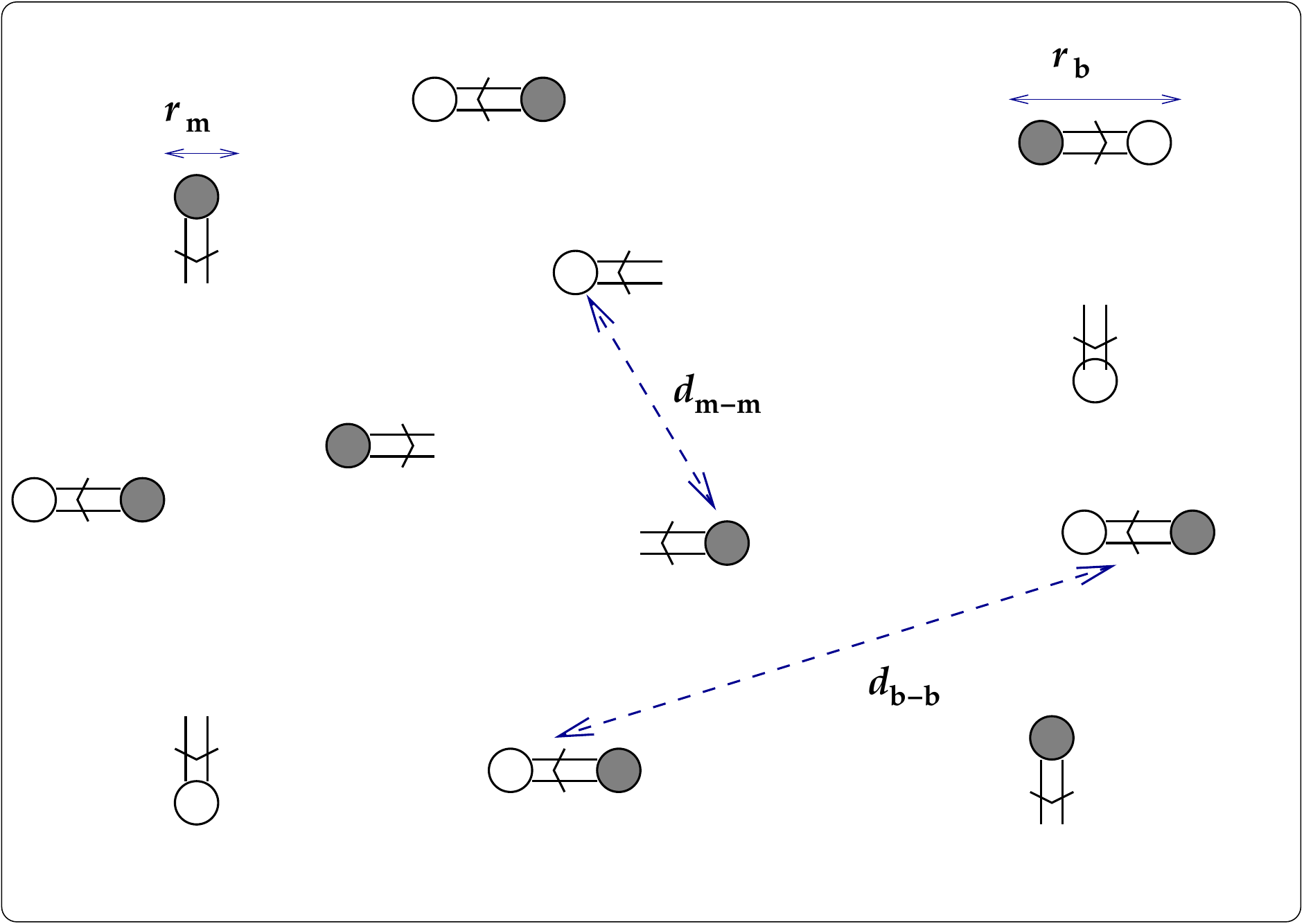}
	\hfil
        \caption
	{\small The Euclidean vacuum of the small-$m,L$ theory 
        can be described as a plasma of monopole-instantons  (gray circles) 
        and anti-monopole-instantons (white circles) with fermionic zero 
        modes (unpaired arrows). The  paired events are magnetic and 
        neutral bions.  Neutral bion amplitudes generate repulsion 
        among the eigenvalues of the Wilson line and magnetic bions 
        generate a mass gap for gauge fluctuations via a generalization 
        of the Polyakov mechanism to a locally 4d theory.}	
        \label{fig:plasma}
        \end{center}
        }
    }
\end{FIGURE}

 Monopole-instantons have two fermionic zero modes associated with 
a vertex of the form  $\frac{\partial^2 W}{\partial \Phi_i \partial \Phi_j} 
\psi_i \psi_j \leftrightarrow \sum_k\cM_k$, where $W(\Phi_i)$ is the 
superpotential of the effective theory.  This means that the superpotential 
can be extracted from the monopole amplitude, but it does not imply that 
the mass gap for bosonic fluctuations is generated by monopoles, as 
is evident from the existence of a fermionic bilinear in the amplitude.    
The bosonic potential can be computed from the superpotential, $V=
\sum_i |\frac{\partial W}{\partial \Phi_i}|^2 \leftrightarrow  \sum_{i, j} 
\cB_{ij}$, but the physical mechanism which generates the bosonic potential 
and the mass gap for bosonic fluctuations is due to bions, correlated 
monopole-anti-monopole instantons without any fermionic zero modes. The 
bosonic potential also has $h^\vee$ minima, leading, at weak coupling,   
to the spontaneous breaking of the discrete chiral symmetry, $\Z_{2 h^\vee} 
\rightarrow \Z_2$. This, in turn, generates a dynamical mass for fermions.    
The importance of this point of view, apart from providing the correct 
interpretation of the physical phenomena  governing the dynamics in the supersymmetric 
theory, is that semi-classical monopole and bion amplitudes also
exist in non-supersymmetric theories, where the bosonic potential 
cannot be extracted from the super-potential \cite{Unsal:2007jx,Anber:2011de,Poppitz:2011wy,Argyres:2012ka}.

\subsection{Phase transition in the small $\mathbf{m}$-$\mathbf{L}$ regime 
and universal aspects} 

There are two main effects of adding a small fermion mass term. The  mass 
term lifts the zero modes of the monopole-instantons. This implies that 
there is a non-zero monopole-instanton contribution to the bosonic potential.  
The mass term also breaks supersymmetry, which leads to a perturbative 
contribution to the potential for the holonomy \cite{Gross:1980br}. Studying 
the competition between these effects and the bion induced potential already 
present at $m=0$ shows that there is a phase transition at some critical 
compactification scale that grows with $m$. We find a  description of 
this phase transition valid for {\it all}  Lie  groups, $G$:

\begin{enumerate}
\item
Neutral bions {\it always} generate repulsion among the eigenvalues 
of the  Wilson line around $\S^1$. For theories with a $\Z_N$ center symmetry, 
the repulsion leads to a $\Z_N$-symmetric distribution, while  for theories 
without a center symmetry, it leads to a  non-degenerate distribution of 
eigenvalues, as we show explicitly for $G_2$.

\item  
Monopole-instantons, along with the perturbative potential for the Wilson 
line, lead to an attractive interaction between the eigenvalues of the 
Wilson line. This interaction increases with $m$. Monopoles prefer 
configurations in which the eigenvalues accumulate. This leads to 
center-instability whenever $ Z(G)$ is non-trivial and to eigenvalue 
bunching whenever $Z(G)$ is trivial.

\item  
Magnetic bions lead to an attraction among the eigenvalues of 
the Wilson line. They also generate a mass gap for gauge fluctuations, 
and they are responsible for the confinement of electric charges.  
However, the combined  effect of neutral and magnetic bions (which are 
of the same order in the semi-classical expansion) always generates a  
repulsion among eigenvalues. 

\item 
The  competition of neutral and magnetic bions on the one hand and
monopole-instanton and perturbative effects on the other hand is 
responsible for the semi-classical realization of deconfinement. 
For all gauge groups but $G_2, F_4, E_8$,  this phase transition 
is associated with a change in the center symmetry realization. 
For $SU(N_c)$ with $N_c\geq 3$,  we find a first order phase transition 
associated with a change in the center symmetry realization.
 In earlier 
work, we found  a second order phase transition for $N_c=2$ 
\cite{Poppitz:2012sw}. In theories without a center symmetry, we find 
a first order transition accompanied by a jump in the Polyakov loop. 
These findings agree with lattice results (see  \cite{Lucini:2005vg, Panero:2009tv, Mykkanen:2012ri}, as well as  \cite{Lucini:2012gg} 
for a recent review), providing support for the conjectured continuity 
of the deconfinement transition depicted in Fig.~\ref{fig:phasediag}.

\end{enumerate}

  Finally, we investigate the topological $\theta$-angle dependence of 
the twisted partition function. As is well-known, turning on a $\theta$-angle  
introduces a sign problem in the Euclidean path integral formulation. 
The semi-classical manifestation of the sign problem is a complex
fugacity of monopole-instantons. The phase of the amplitude generates 
interference  between Euclidean path histories, which can be interpreted 
as the analytic continuation of Aharonov-Bohm type interference.
We refer to this phenomenon as  ``topological interference". In the 
case of finite rank gauge groups topological interference leads to
$\theta$ dependence of the critical temperature. The minimum $T_c$
occurs at $\theta=\pi$. At this point the $\theta$ dependence is 
non-analytic, and all theories have two degenerate vacua in the 
confined phase.  At $N=\infty$ the critical temperature is independent
of $\theta$.

\section{$\mathbf{SU(N_c)}$,  $\mathbf{N_c \geq 3}$: First order  
center symmetry changing phase transition }

As reviewed in Section~\ref{sec:review}, at sufficiently small $L$ 
the  supersymmetric $SU(N_c)$ theory on $\R^3 \times \S^1$ dynamically 
abelianizes down to  $U(1)^{N_c -1}$ at distances larger than $LN_c/ 2\pi$.  
In this regime the light bosonic fields are the $N_c-1$  dual photon 
fields $\vec{\sigma}$ (a $N_c-1$-dimensional vector) and their scalar 
superpartners $\vec{\phi}$, which are the ``uneaten" components of 
the gauge connection along $\S^1$. It is convenient to expand the  
fields $\vec\phi$ and $\vec\sigma$  as follows:
\begin{eqnarray}
\label{fields}
\vec\phi  &\equiv& {2 \pi \over {N_c}}\; \vec\rho 
    + {g^2 \over 4 \pi} \;\vec{b}^\prime , \nonumber \\
\vec\sigma + {\theta \vec{\phi} \over 2 \pi} &\equiv& 
          {2 \pi k + \theta \over {N_c}} \; \vec\rho 
      + \vec\sigma^\prime~.
\end{eqnarray}
The primed fields describe the fluctuations of the fields $\vec{\phi}$ 
and $\vec{\sigma}$ around their supersymmetric and center-symmetric 
ground state. For every $N_c$ the supersymmetric ground state is labeled 
by an integer $k = 1,\ldots N_c$, the vacuum angle $\theta$, and the Weyl 
vector $\vec{\rho}$. The integer $k$ corresponds to a choice of vacuum 
corresponding to the spontaneous breaking of the anomaly-free discrete 
chiral ($R$-) symmetry, $\Z_{2 N_c} \rightarrow \Z_2$. 

We normalize the  $SU(N_c)$ Cartan generators $\vec{H} = (H_1,\ldots,
H_{N_c - 1})$ by $\tr [H_a H_b] = \delta_{ab}$. All the $SU(N_c)$ roots 
have length squared equal to 2, and the roots and co-roots, $\vec\alpha$ 
and $\vec\alpha^*$, as well as the fundamental weights and co-weights, 
$\vec\omega$ and $\vec\omega^*$, are the same. The Weyl vector is given 
by $\vec\rho = \sum_i \vec\omega_i$. The fundamental weights obey 
$\vec\omega_{i} \cdot \vec\alpha_j = \delta_{ij}$ ($i,j=1, \ldots N_c - 1$), 
and the Weyl vector obeys $\vec\rho \cdot \vec\alpha_{i} = 1$, for $i = 1, 
\ldots N_c - 1$, and $\vec\rho \cdot \vec\alpha_{N_c} = 1 - N_c$.
 
 The fields $\vec{b}'$ and $\vec{\sigma}'$ in (\ref{fields}) are periodic,
\begin{equation}
\label{periodicities}
\vec{b}^\prime \sim \vec{b}^\prime + {8 \pi^2 \over g^2} \; 
           \vec{\omega}\, ,  \quad
\vec\sigma^\prime \sim \vec\sigma^\prime 
      + 2 \pi \vec\omega\, . 
\end{equation}
Note that for small $g$ periodicity of $\vec{b}^\prime$ is irrelevant.  
The gauge holonomy (Wilson loop  around $\S^1_L$) in terms of the fields 
(\ref{fields}) is given by:
\begin{equation}
\label{wilsonloop}
\Omega = \exp \left({ i\; {2 \pi \over N_c} \vec{H} \cdot \vec\rho 
  + i\; {g^2 \over 4 \pi} \vec{H} \cdot \vec{b}^\prime }\right)~.
\end{equation}
At $\vec{b}^\prime = 0$, $\Omega$ obeys $\tr\;\Omega^n = 0,  \forall  \; n 
\neq 0 \;({\rm mod} \; N_c)$. We note that $\tr\;\Omega^{n}$ are order 
parameters of the $\Z_{N_c}$ global center symmetry that  the adjoint-fermion 
$SU(N_c)$ theory acquires when compactified on $\R^3 \times \S^1_L$ and that 
center symmetry is unbroken in any of the $N_c$ supersymmetric vacua.  
  
  It is also important to note that the Wilson line that satisfies  
$\langle \tr\;\Omega^n \rangle= 0$ at weak coupling is associated with 
the eigenvalue configuration depicted on Figure~\ref{fig:phase2}(b). We 
refer to this type of gauge holonomy as weak coupling non-trivial holonomy. 
At strong coupling, there is no adjoint Higgsing, as the fluctuations of 
the eigenvalues are larger than the typical inter-eigenvalue separation, 
hence in the latter, the eigenvalues are essentially randomized over the 
dual circle on which the eigenvalues live. We will discuss this distinction
in more detail in Section~\ref{sec:weakstrong}. 

\begin{FIGURE}[ht]{
    \parbox[c]{\textwidth}
        {
        \begin{center}
        \includegraphics[angle=0, scale=0.4]{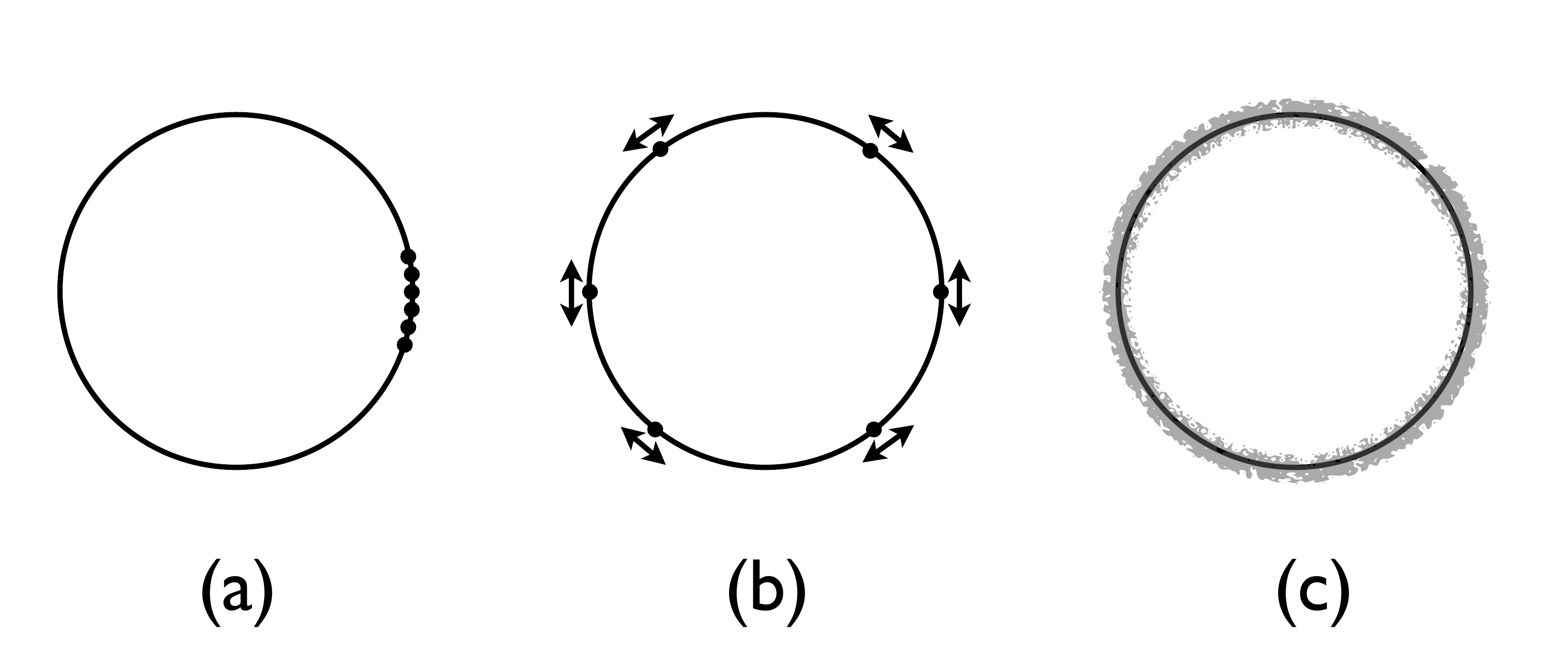}
	\hfil
        \caption
	{\small Three types of gauge holonomy: 
	{\bf a}): Weak coupling trivial holonomy associated with deconfined 
        phase. 
	{\bf b}): Weak coupling non-trivial holonomy associated with the weak 
        coupling confined phase.
	{\bf c}):  Strong  coupling non-trivial holonomy associated with 
        the strong coupling confined phase.
        It is important to note that {\bf b}) admits an adjoint-Higgs 
        description, whereas {\bf c}) does not. These two confinement 
        regimes are continuously connected. }	
        \label{fig:phase2}
        \end{center}
        }
    }
\end{FIGURE}

\subsection{Relevant scales and two large-$\mathbf{N_c}$ limits:  
't Hooft vs. Abelian}
\label{section:large-N-lim}

 The gauge theory compactified on $\R^3\times \S^1$ contains an extra 
parameter, the compactification scale $L$. Thus, if center symmetry remains 
unbroken at small $L$, the large-$N_c$ limit 
is more subtle than the standard 't Hooft limit.\footnote{This issue does not arise in thermal 
compactifications, and as a consequence it has not been discussed until
recently, see  \cite{Unsal:2010qh}.}  In fact, there are two large-$N_c$ 
limits, which we refer to as the (naive) 't Hooft limit and the abelian 
limit. The reason that there is more than one limit is that $m_W$, the 
mass of the lightest off-diagonal gluon, goes to zero if the large-$N_c$ 
limit is taken with $L\Lambda$ fixed. This means that the effective abelian 
theory breaks down in the naive 't Hooft limit. When we study the large-$N_c$ 
limit we will therefore consider the limit $N_c\to\infty$ with $m_W = 
2 \pi/(N_cL) = {\it const}$, i.e., $L=O(N_c^{-1})$. In other words, ensuring 
the validity of the weakly-coupled effective theory requires shrinking the 
size of the circle when taking the large-$N_c$ limit. In what follows, we 
will express all dimensionful quantities in the effective lagrangian for 
the light modes on $\R^3 \times \S^1$ in terms of $m_W$ and the scale 
parameter of the gauge theory, $\Lambda$, with the understanding that 
they are both fixed  in the large-$N_c$ limit,
\begin{eqnarray}
\label{parameters}
\Lambda^3 &=& \mu^3 \; \frac{16\pi^2}{3N_cg^2(\mu)} 
  \exp\left( -\frac{8\pi^2}{g^2(\mu) N_c}\right) ~,\nonumber \\
  m_W &=& {2 \pi \over N_c L}  \gg \Lambda~. 
\end{eqnarray}
The above expression for $\Lambda$ is the standard 2-loop expression 
for the scale parameter \cite{Beringer:1900zz}. Note that 
Ref.~\cite{Davies:2000nw} uses a definition of $\Lambda^3$ that 
differs by a factor of proportional to $N_c$. The gaugino mass $m$ 
introduces another length scale in the problem. The validity of the 
fermion zero mode induced pairing mechanism of bions requires that 
$m \ll g^2(m_W)/L$, the inverse bion size. In terms of the scales in  
Eq.~(\ref{parameters}), the gaugino mass should parametrically obey
\begin{equation}
\label{gauginolimit}
m \ll   {m_W \over \log {m_W\over \Lambda}}~,
\end{equation}
an upper bound which remains finite in the large-$N_c$ limit.

\subsection{The bion and monopole potentials in softly-broken 
$\mathbf{SU(N_c)}$ theory on $\mathbf{\R^3 \times \S^1}$}
\label{sec:Veffm}

Here, we describe the Lagrangian governing the long-distance dynamics of 
the  $SU(N_c)$ gauge theories on $\R^3 \times \S^1_L$ for general $N_c$. 
The effective Lagrangian follows from the superpotential of 
Ref.~\cite{Davies:2000nw} and the field redefinition in Eq.~(\ref{fields}) 
in a rather straightforward way, generalizing the detailed derivation
given in \cite{Poppitz:2012sw} for $N_c=2$. The kinetic term of the light 
fields (\ref{fields}) is:\footnote{
The kinetic terms of $\vec{b}^\prime$ and $\vec{\sigma}'$ receive 
perturbative corrections along the Coulomb branch. These are closely  
related to the non-cancelling one-loop fluctuation determinants 
around  supersymmetric instanton-monopole solutions on $\R^3 \times 
\S^1$. This subtlety is studied in detail in Appendix A of 
\cite{Poppitz:2012sw}. However, as in that reference, our results in 
the weakly-coupled calculable regime are unaffected to leading order.}
\begin{equation}
\label{kinetic1}
{\cal L}_{kin.} = {g^2(m_W) \over 16 \pi^2 L }
    \left( (\partial_\mu \vec{b}^\prime)^2 
         + (\partial_\mu \vec\sigma^\prime)^2 \right)\, , 
\end{equation}
where $m_W$ is given in (\ref{parameters}). The nonperturbative potential 
due to bions and monopole-instantons is:
\begin{eqnarray}
\label{totalv}
\label{V_np_Nc}
V_{np}^{(k)}  &=& ~~ V^0_{\it bion} \left[ \sum\limits_{i = 1}^{N_c} 
     e^{ - 2 \vec\alpha_i \cdot \vec{b}^\prime} 
   - e^{- ( \vec\alpha_i + \vec\alpha_{i+1}) \cdot \vec{b}^\prime } 
        \cos \left[(\vec\alpha_i - \vec\alpha_{i+1}) 
           \cdot \vec\sigma^\prime \right] \right]  \\
&& - V^0_{\it mon} 
    \left[ \sum\limits_{i = 1}^{N_c} 
       \left( 1 +  {g^2 N_c \over 8\pi^2}
                \vec{\alpha}_i \cdot \vec{b}^\prime \right)  
        e^{  -  \vec\alpha_i \cdot \vec{b}^\prime}        
        \cos\left( \vec\alpha_i  \cdot \vec\sigma^\prime 
               + {2 \pi k + \theta \over N_c}
\right)   \right]~,\nonumber 
\end{eqnarray}
where we have included the dependence on the $\theta$ parameter of 
the gauge group, which only contributes if the gaugino mass is nonzero. 
The parameter $k=0, 1,...,N_c-1$ labels the vacuum branch as in 
Eq.~(\ref{fields}). The true effective potential corresponds to the branch 
that gives the minimum ground state energy, the other branches describe
metastable vacua. These vacua are expected to become quasi-stable in
the large-$N_c$ limit \cite{Witten:1998uka}. The strengths of the bion 
and monopole potentials are given by
\begin{eqnarray}
\label{bion1}
V^0_{\it bion} &=& \frac{32 \pi^2 L}{g^2(m_W)}
 \frac{\mu^6 L^2}{g^4(\mu)} 
\exp\left(-\frac{16\pi^2}{g^2(\mu) N_c} \right)\, , \\
\label{bion2}
V^0_{\it mon} &=& \frac{4Lm\mu^3}{g^2(\mu)}
     {8\pi^2 \over g^2(\mu) N_c}
  \exp\left(-\frac{8\pi^2}{g^2(\mu) N_c} \right)\, . 
\end{eqnarray}
For $N_c=2$ and $\theta = 0$ the above result reduces to the expressions 
given in  \cite{Poppitz:2012sw}. We can express these results in terms
of the physical scales defined in Eq.~(\ref{parameters}). The kinetic
term is 
\begin{equation}
\label{kinetic2}
{\cal L}_{kin.} = {1 \over 12 \pi} {m_W \over \log(\frac{m_W}{\Lambda})}
    \left( (\partial_\mu \vec{b}^\prime)^2 
         + (\partial_\mu \vec\sigma^\prime)^2 \right)\, , 
\end{equation}
and the coefficients  (\ref{bion1}, \ref{bion2}) of the bion and monopole 
potentials are
\begin{eqnarray}
\label{bion3}
V^0_{\it bion} &=& \frac{27}{8\pi}  
  \frac{\Lambda^6}{m_W^3} \log\left(\frac{m_W}{\Lambda}\right) \,  , \\
  \label{bion4}
V^0_{\it mon} &=&  \frac{9}{2\pi}
  \frac{m\Lambda^3}{m_W} \log\left(\frac{m_W}{\Lambda}\right) \, . 
\end{eqnarray}
For future use, we also introduce the dimensionless ratio of Eq.~(\ref{bion4}) 
and (\ref{bion3}):
\begin{equation}
\label{cm}
 c_m= \frac{V^0_{\it mon}}{V^0_{\it bion}} 
  = \frac{4mm_W^2}{3\Lambda^3} 
  = \frac{16 \pi^2 m }{3\Lambda  (\Lambda L N_c)^2}  \, . 
\end{equation}
We first study the spectrum in the limit when the supersymmetry breaking 
perturbation is turned off, $m=0$. For this purpose we expand the bion 
potential to quadratic order in $\vec{b}'$ and $\vec\sigma'$. Because 
of the exact supersymmetry the $\vec{b}'$ and $\vec\sigma'$ masses are 
identical and we can concentrate on the $\vec{b}'$ field. We rescale 
$\vec{b}'$ to canonically normalize the kinetic term (\ref{kinetic2}), 
and also drop the prime in the following. We find the following quadratic 
lagrangian
\begin{equation}
\label{masssusy}
{\cal L} = \frac{1}{2}(\partial_\mu \vec{b})^2
 + m_0^2 \sum_{i=1}^{N_c} \left[ 
  \left(\vec\alpha_i\cdot \vec{b}\right)^2 
  -  \left(\vec\alpha_i\cdot \vec{b}\right)
       \left(\vec\alpha_{i+1}\cdot \vec{b}\right)\right]\, .
\end{equation}
with
\begin{equation}
m_0^2 = \frac{81}{4}
 \frac{\Lambda^6\left[\log(\frac{m_W}{\Lambda})\right]^2}{m_W^4}\, . 
\end{equation}
The mass term in (\ref{masssusy}) can be diagonalized by switching 
to $U(N_c)$ roots given by $N_c$-dimensional vectors of the form 
$\alpha_1= (1,-1,0,\ldots)$, $\alpha_2=  (0,1,-1,0,\ldots)$, etc., 
$\alpha_{N_c}$ $=$ $(-1,0, \ldots, 0,1)$, see \cite{Unsal:2008ch}. 
This introduces an extra massless particle that decouples from the 
rest of the spectrum. We get
\begin{equation}
{\cal L} = \frac{1}{2}(\partial_\mu \vec{b})^2
 + \frac{1}{2} \;m_0^2 \sum_{i=1}^{N_c} \left( 
   b_{i+2}-2b_{i+1}+b_i \right)^2\, .
\end{equation}
This is a simple lattice model that can be diagonalized by introducing 
$\Z_{N_c}$ Fourier modes
\begin{equation}
\label{fourier}
 b_j= \frac{1}{\sqrt{N_c}} \sum_{p=0}^{N_c-1} \tilde{b}_p
  e^{-2\pi i\frac{pj}{N_c}}\, .
\end{equation}
The spectrum of the $\vec{b}$ fields is 
\begin{equation}
  m_b =  4 m_0   \sin^2\left( \frac{p\pi}{N_c} \right)  
      = 18 \Lambda \; {\Lambda^2  \log {m_W \over \Lambda}\over m_W^2}  
    \sin^2\left( \frac{p\pi}{N_c} \right)~, 
    \qquad (p=1,\ldots,N_c-1)\, ,  
\label{mb}
\end{equation}
where we have dropped the unphysical massless mode $p=0$ (which was 
introduced above via the transition to $N_c$-dimensional roots). From 
Eq.~(\ref{mb}) we conclude that all the $\vec{b}$-fields are parametrically  
lighter than the $W$-bosons and that the effective theory is consistent in 
the large-$N_c$ limit, see Eq.~(\ref{parameters}). 

Let us now turn on the gaugino mass $m$. At first, we  ignore the 
$\theta$-dependence and  take $\theta = 0$ and $k=0$ (it is easy to 
see that for $\theta=0$ the $k=0$ vacuum in Eq.~(\ref{fields}) has the
minimal energy). In this vacuum, there is no mixing between the 
$\vec{b}'$ and $\vec{\sigma}'$ fields when the monopole potential 
is expanded up to   quadratic terms around their vanishing expectation 
values. In the weak coupling limit ${g^2N_c\over 8\pi^2} = [3 \log(m_W/
\Lambda)]^{-1}$, so we can neglect the $g^2N_c$ term in the monopole 
potential in  Eq.~(\ref{V_np_Nc}). We now find, instead of (\ref{masssusy}), 
the quadratic Lagrangian
\begin{equation}\label{mass2}
{\cal L} = \frac{1}{2}(\partial_\mu \vec{b})^2
 + m_0^2 \sum_{i=1}^{N_c} \left[ 
      \left(\vec\alpha_i\cdot \vec{b}\right)^2 
    - \left(\vec\alpha_i\cdot \vec{b}\right)
        \left(\vec\alpha_{i+1}\cdot \vec{b}\right) 
    - {c_m \over 2} (\vec{\alpha}_i \cdot \vec{b})^2\right]\, .
\end{equation}
Thus, the monopole term is diagonal in the Fourier basis given in
Eq.~(\ref{fourier}), and the only difference is that the quadratic part 
of the monopole potential has eigenvalues that are proportional to the 
square of $\sin(p\pi/N_c)$.  The squared mass of the $\vec{b}$ field as 
a function of $m$ is
\begin{equation}
\label{mbmass2}
  m_b^{2} = 16 m_0^2 \left\{
    \left[\sin\left(\frac{p\pi}{N_c}\right)\right]^4
  - { c_m \over 4} \left[\sin\left(\frac{p\pi}{N_c}\right)\right]^2 
   \right\}\, . 
\end{equation}
We observe that the center symmetric vacuum becomes locally unstable 
for $c_m>c_m^{*}$, where, taking $p=1$, $c_m^{*} = 4 \sin^2(\frac{\pi}
{N_c})$. This corresponds to $m>m^{*}$ with 
\begin{equation}
\label{m_cr_loc}
  m^{*} =  \frac{3 \Lambda^3}{m_W^2} \sin^2 \left(\pi \over N_c\right)
     \stackrel{N_c \rightarrow \infty}{\longrightarrow} 
     \frac{3 \pi^2 \Lambda^3}{N_c^2m_W^2} ~.
\end{equation}
for fixed $L$ as one varies $m$. On the other hand, for fixed $m$, as 
one varies $L$, this implies a local instability at 
\begin{equation}
\label{L_cr_loc}
L^* = \Lambda^{-1} \sqrt {\frac{4m}{3 \Lambda } } 
 \frac{1}{\frac{N_c}{\pi} \sin \frac{\pi}{N_c}}
     \stackrel{N_c \rightarrow \infty}{\longrightarrow} 
   \Lambda^{-1} \sqrt {\frac{4m}{3 \Lambda}}~.
\end{equation}
It is clear that $m^{*}$ (or $L^*$)  is within the region of validity
of the semi-classical approximation given in Eq.~(\ref{gauginolimit}).

 We note, however, that for any $N_c>2$ the non-perturbative potential 
contains cubic terms. This implies that the transition is first order,
and that $c_{m}^{*}$ is not the critical coupling, but rather it is the 
limit of metastability above which the confining vacuum ceases to be 
a (local or a global) minimum. The critical mass, $m_{\it cr}$, is smaller
than $m^{*}$  in Eq.~(\ref{m_cr_loc}). We will determine the critical 
values of $c_m^{\it cr}$ in Section \ref{su3transition}. For $c_m^{\it cr} 
< c_m<c_m^{*}$ the center-symmetric minimum is a metastable local minimum 
but not the global minimum. 

\subsection{The perturbative potential for arbitrary $\mathbf{N_c}$}   
\label{sec:pert}

 Before we continue our study of the phase transition we will first 
show that the perturbative (Gross-Pisarski-Yaffe) contribution to 
the potential for the holonomy is sub-leading compared to the monopole 
and bion induced potentials, as in the $N_c =2$ case. The perturbative 
potential for the Polyakov line in $SU(N_c$) is
\begin{equation}
V_{\it pert} = -\frac{m^2}{L} \sum_{N_c\ge j>i\ge1} 
  B_2\left( \frac{q_{ij}}{2}\right)\, , 
\end{equation}
where we have neglected higher order terms in the mass of the 
adjoint fermion. Here, $B_2(x)=x^2 - x + 1/6$ is the Bernoulli polynomial 
of order 2, and
\begin{equation}
 q_{ij} = \frac{2}{N_c}(j-i) + \frac{g^2}{4\pi}
  \left(b_i'-b_j'\right) \, .
\end{equation}
For small $b_i'$ we can ignore the periodicity of the potential.
The center symmetric point $b_i'=0$ is a local  {\it maximum} of the 
perturbative one-loop potential. At order $m^2$ the potential near 
the center symmetric vacuum is exactly quadratic, 
\begin{equation}
 V_{\it pert} =-\frac{m_{\it pert}^2}{2N_c} 
     \sum_{j>i} \left(b_i-b_j\right)^2 \, , 
 \hspace{0.5cm} 
  m_{\it pert}^2 = \frac{2m^2 \pi^2}{3\log(\frac{m_W}{\Lambda})}\, , 
\end{equation}
where we have dropped a constant term and normalized the kinetic term  
as in Eq.~(\ref{masssusy}). The sum can be diagonalized in terms of the 
$\Z_{N_c}$ Fourier modes given in Eq.~(\ref{fourier}). The tachyonic mass 
for the $p$'th mode is
\begin{equation} 
 m_b^2 = -\frac{m_{\it pert}^2}{4N_c} \sum_{j>i} 
 \left[ \sin\left( \frac{\pi p (j-i)}{N_c}\right)\right]^2
 = -  m_{\it pert}^2\, , 
\end{equation} 
independent of $p$. We can compare this result to bion generated 
mass of the lowest mode, 
\begin{equation} 
 m_b^2 \sim \frac{\Lambda^6}{N_c^4m_W^4} 
 \left[\log\left(\frac{m_W}{\Lambda}\right)\right]^2\, . 
\end{equation}
At the critical mass $m^{*}$ defined in equ.~(\ref{m_cr_loc})
the tachyonic mass is 
\begin{equation} 
 m_b^2 \sim \frac{\Lambda^6}{N_c^4m_W^4} 
  \left[\log\left(\frac{m_W}{\Lambda}\right)\right]^{-1}\, ,
\end{equation}
and we conclude that the perturbative contribution is suppressed by 
three powers of $\log(m_W/\Lambda)$. This is the same suppression 
factor we found in $N_c=2$ \cite{Poppitz:2012sw}. We will henceforth 
ignore the contribution of the perturbative fluctuations to the 
potential for the holonomy.

\begin{figure}[t]
    {
    \parbox[c]{\textwidth}
        {
	\begin{center}
        \includegraphics[angle=0, width=0.45\textwidth]{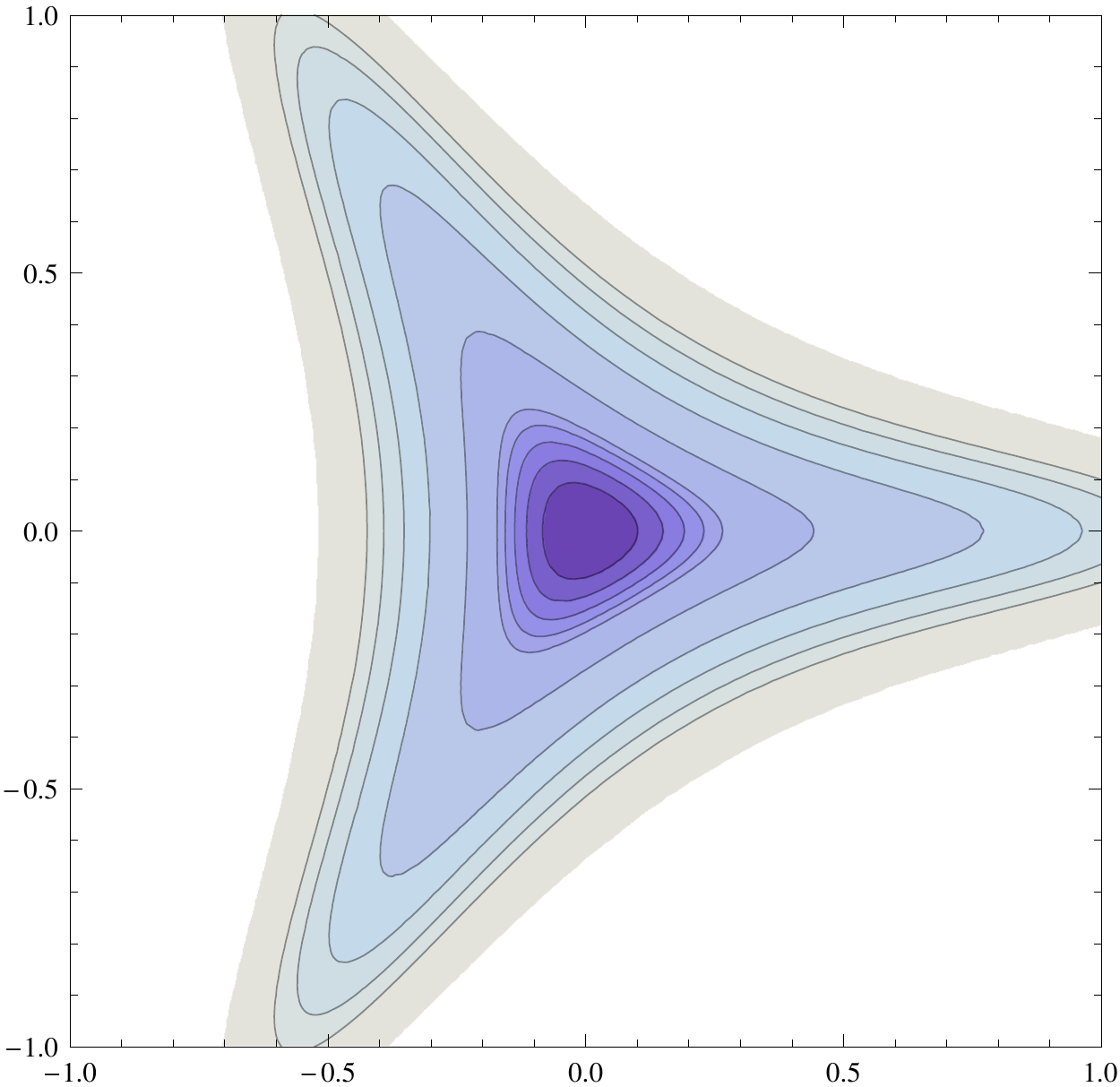}
        \hspace*{0.05\textwidth}
        \includegraphics[angle=0, width=0.45\textwidth]{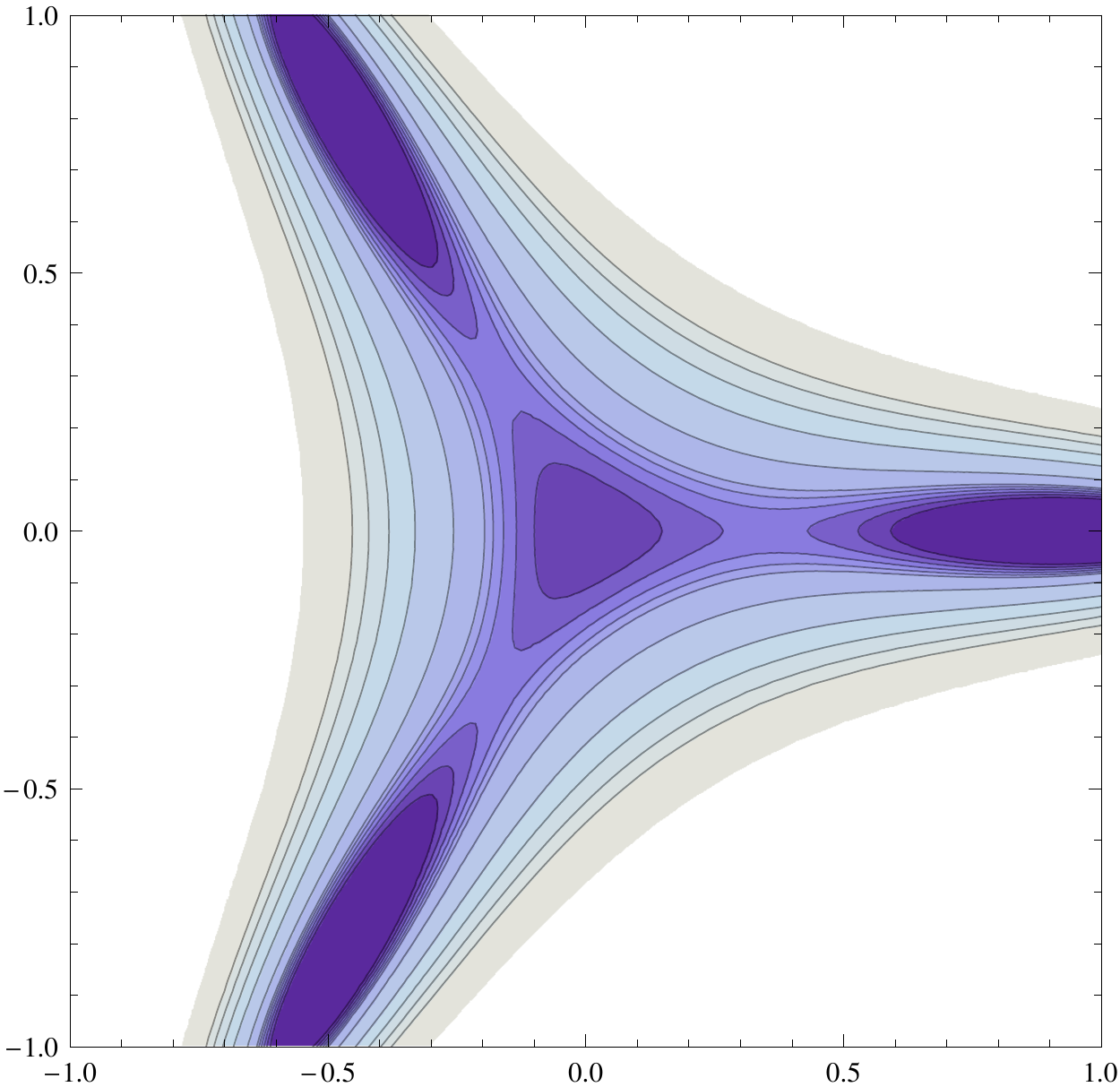}
        \end{center}
	\caption{\small 
        Contour plots of the bion- and monopole-instanton-induced potential, 
        as a function of the two holonomies, showing the first order phase 
        transition for $SU(3)$ (darker shades represent smaller values of 
        the potential). Left panel: Contour plot for $c_m < c_m^{**}<c_m^{\it cr}$ 
        ($c_m = 2.20$, $c_m^{\it cr} = 2.446$) as a function of $b_1, b_2$. 
        The $\Z_3$-symmetric  (confining) minimum is at the origin. 
        Right panel: Contour plot for $ c_m^{\it cr} < c_m < c_m^{*} $ 
        ($c_m =2.5$) as a function of $b_1, b_2$. The $\Z_3$-breaking global 
        minima are clearly visible, and the  $\Z_3$-symmetric confining minimum is meta-stable.}
        \label{fig:su3}
        }
      }
\end{figure}

\subsection{The phase transition  and the limits of metastability for 
$\mathbf{N_c \ge 3}$}
\label{su3transition}

 In Section \ref{sec:Veffm} we determined the limit of metastability 
$c_m^{*}$ at which the origin of the effective potential for $N_c\geq 3$ 
becomes unstable, and the metastable phase disappears. In this section we 
will numerically determine the critical $c_m^{\it cr}$ for the first 
order phase transition. 

\begin{figure}[t]
	\begin{center}
        \includegraphics[angle=0, width=0.65\textwidth]{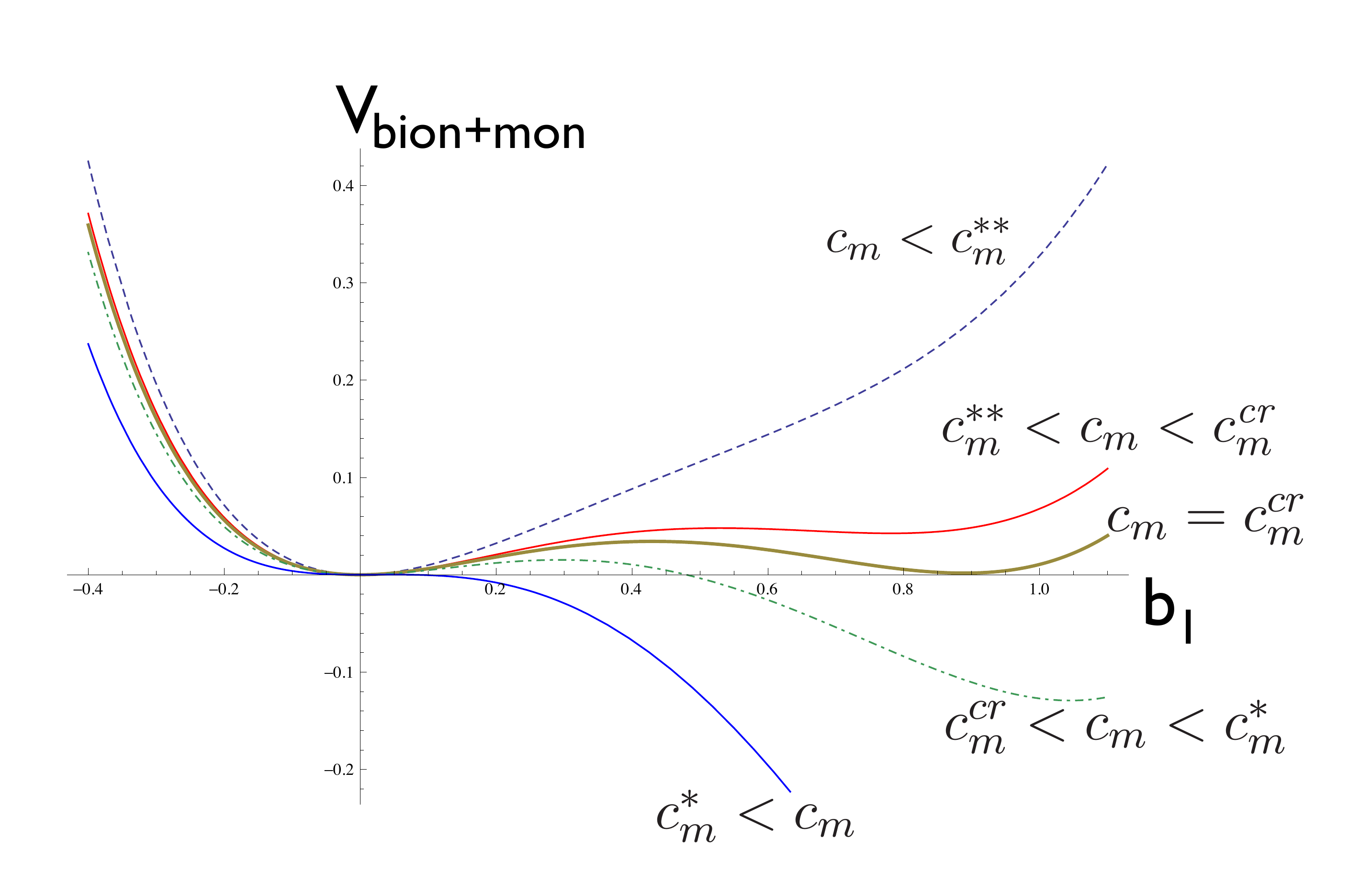}
        \end{center}
        \caption{\small
        Effective potential due to bions and monopoles for different
        values of $c_m$. The potential (in units of $V_{bion}^0$) is shown
        as a function of $b_1$ for $b_2 = 0$, corresponding to a cut
        along the $x$-axis in 
        Fig.~4.          
        \label{fig:su3b}  }
\end{figure}

{\flushleft{{\bf SU(3):}}} For $N_c=3$, one can plot the potential 
as a function of the deviation of the holonomies from their center 
symmetric values (i.e., $b_1', b_2'$) for different values of $c_m$ 
and observe the nature of the transition, see Figure \ref{fig:su3}.
For $\theta = 0$, the expectation value of $\vec{\sigma}'$ remains 
zero. The critical value for the first order transition is $c_m^{\it cr}
=2.446$, see Figure \ref{fig:su3b}. 

 For $c_m < c_m^{**}$, where $c_m^{**}$ is one of the  limits of metastability, 
there  is a unique center-symmetric minimum  (confining phase), see the 
left panel of Figure \ref{fig:su3} as well as Figure \ref{fig:su3b}.   
For $c_m^{**}  < c_m< c_m^{\it cr}$, there is a center-symmetric global 
minimum (confined phase)  and three center-broken metastable (deconfined) 
local minima (not shown in Figure \ref{fig:su3},  but shown in  
Figure \ref{fig:su3b}).  This also provides an alternative definition 
$c_m^{**}$. It is the value of $c_m$ below which  the metastable 
center-broken minima disappears, hence the name limit of meta-stability.  

For $c_m^{\it cr} < c_m < c_m^{*}$, where $c_m^{*}=3$ is the limit of 
metastability, the three $\Z_3$ breaking minima are global degenerate 
minima, and the center-symmetric confining phase is a local but not a 
global minimum, see the right panel of Figure~\ref{fig:su3}. In this 
regime the confining phase is meta-stable. Finally, for  $c_m > c_m^{*}$, 
the center-symmetric point ceases to be a local minimum, and this 
correspond to the other limit of metastability. This case is not shown 
in Figure \ref{fig:su3},  but shown in  Figure \ref{fig:su3b}.  

{\flushleft{{$\bf SU(N_c), N_c >3$:}}}   The general structure that emerges 
for $SU(N_c), N_c >3$ is similar to the $SU(3)$ case shown in  Figure
\ref{fig:su3b}. We have four characteristic domains for the bion and 
monopole-instanton induced potential:  

\begin{itemize} 
\item $c_m < c_m^{**}$ or $L >L^{**}$:  
There is a unique center-symmetric (confined) minimum.

\item $c_m^{**} < c_m < c_m^{\it cr}$ or $L^{**}>L>L^{\it cr}$: 
A global center-symmetric (confined) minimum and $N_c$ meta-stable 
$\Z_{N_c}$ breaking (deconfined) minima.

\item $c_m^{\it cr} < c_m < c_m^{*}$ or $L^{\it cr}>L>L^{*}$:  
A metastable center-symmetric  (confined) minimum and $N_c$  
global $\Z_{N_c}$ breaking (deconfined) minima. 

\item $c_m^{*} < c_m $ or $L^{*} > L$: $N_c$ center-breaking  global  
(deconfined) minima. 
\end{itemize}

 A lattice study of the endpoint of the regime of metastability in pure 
Yang-Mills theory was reported in \cite{Bringoltz:2005xx}. This study was 
motivated by the old idea that in large-$N_c$ QCD the Hagedorn temperature 
$T_H$, the limiting temperature of a hadronic resonance gas, must correspond 
to a second order phase transition which occurs above the (first order) 
deconfining transition \cite{Cabibbo:1975ig,Thorn:1980iv,Cohen:2006qd}.
The regime $T_c<T<T_H$ is therefore metastable. This behavior was also
observed in large-$N_c$ theories compactified on on $S^3 \times S^1$
\cite{Aharony:2003sx,Aharony:2005bq}. 

\begin{figure}[t]
    {
    \parbox[c]{\textwidth}
        {
	\centering
	\begin{subfigure}{0.32\textwidth}
        \begin{center}
        \includegraphics[angle=0, width=0.75\textwidth]{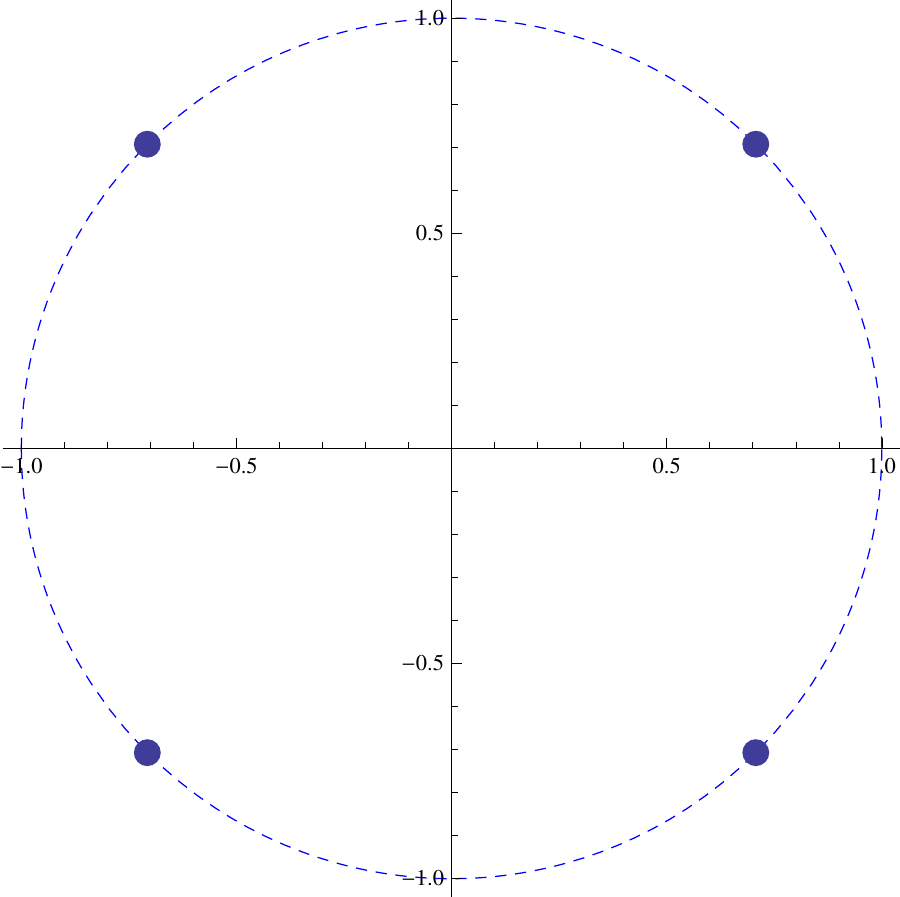}
        \end{center}
	\caption{$c_m \le c_m^{\it cr}$}
	\end{subfigure}
	\centering
	\begin{subfigure}{0.32\textwidth}
        \begin{center}
        \includegraphics[angle=0, width=0.75\textwidth]{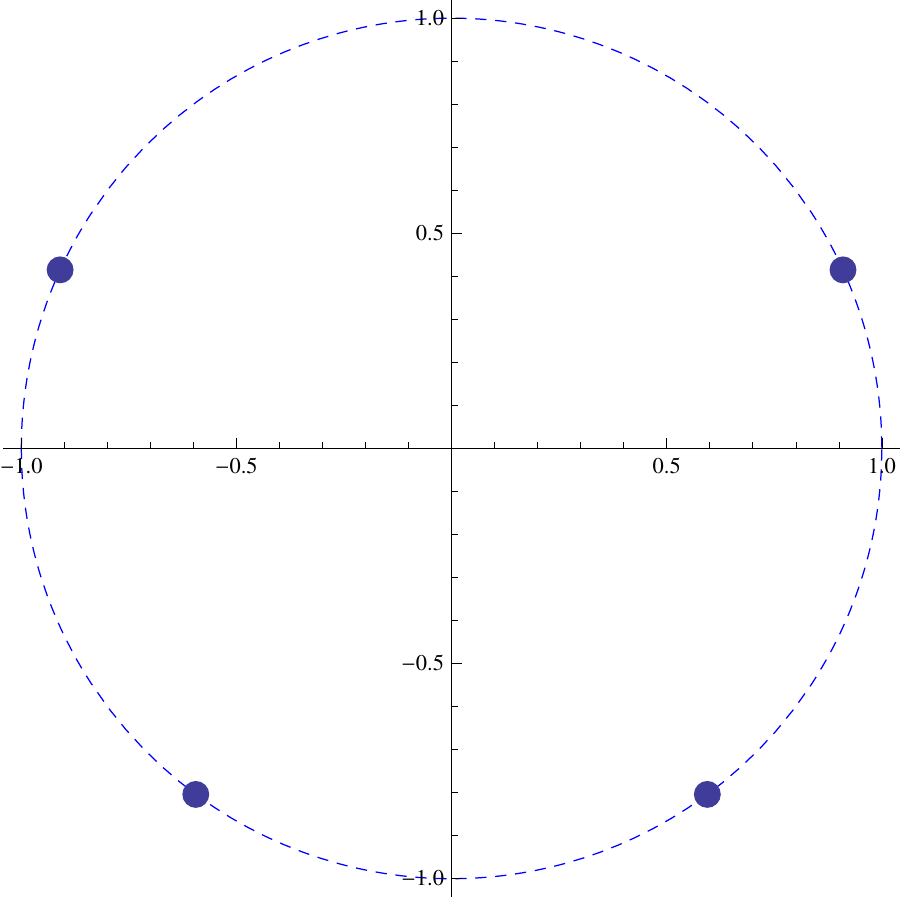}
	\caption{$c_m = c_m^{\it cr} \approx 1.47$ }
        \end{center}
	\end{subfigure}
	\centering
	\begin{subfigure}{0.32\textwidth}
        \begin{center}
        \includegraphics[angle=0, width=0.75\textwidth]{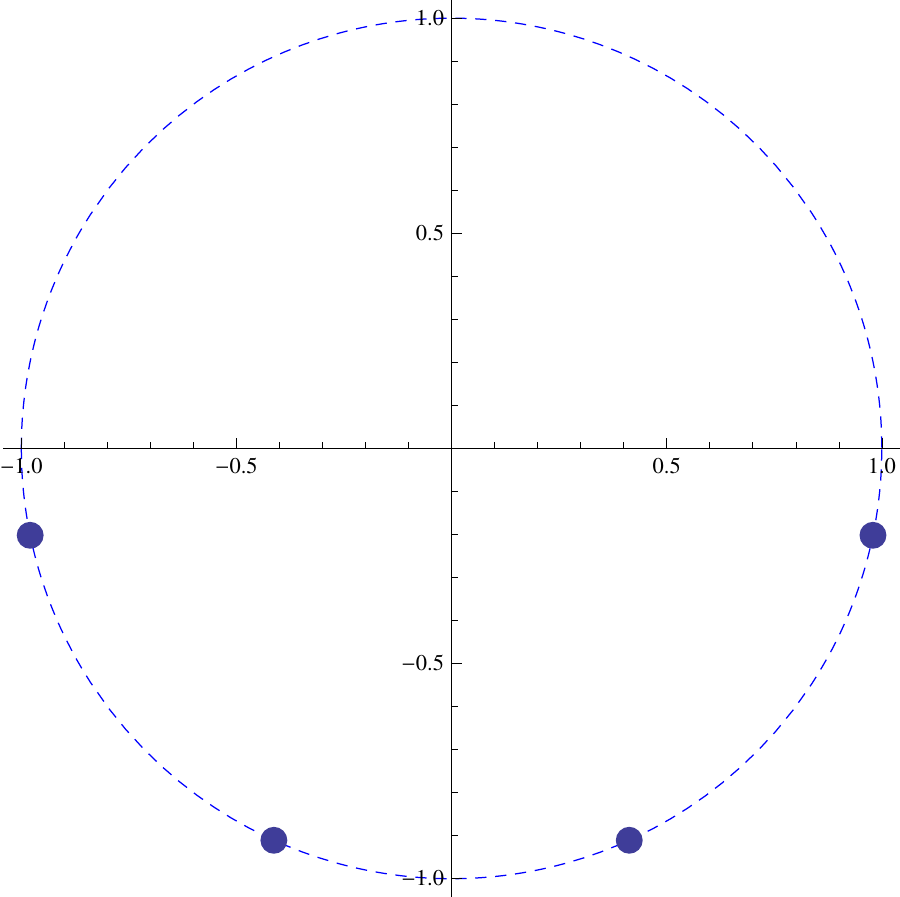}
        \end{center}
	\caption{$c_m = 2 c_m^{\it cr}$  }
	\end{subfigure}
	\hfil
        \caption{\small 
        The distribution of the eigenvalues of the Polyakov loop 
        around $\S^1_L$ for $N_c = 4$ for different values of $c_m$, 
        shown for $g^2 N_c = 0.1$.}
        \label{fig:su4}
        }
      }
\end{figure}

For $N_c >3$ the potential is a function of $N_c-1$ holonomies. Instead of
trying to plot this function we will study the vacuum expectation values 
of the $N_c$ eigenvalues of the Wilson line around $\S^1$ as a function of 
the gaugino mass parameter $c_m$. We will, again, observe a discontinuous 
transition, which occurs at values of $c_m^{\it cr}$ smaller than the  
$c_m^*$  given in  (\ref{m_cr_loc}). The eigenvalue repulsion due to the 
neutral (center-stabilizing) bions is dominant at small $c_m$ and is, 
upon increasing $c_m$, countered by the eigenvalue attraction due to 
monopole-instantons along with the one-loop potential which is  sub-leading,   
see Section~\ref{sec:pert}. We plot the eigenvalues of $\Omega$ as 
$c_m^{\it cr}$ is crossed.  We show the behavior of the eigenvalue 
distributions for $N_c =4,5$, and $10$ in Figures \ref{fig:su4}, 
\ref{fig:su5}, and \ref{fig:su10}, respectively. In order to make these 
plots we chose a specific value of the 't Hooft coupling, $g^2N_c=0.1$. 
The critical value $c_m^{\it cr}$ is independent of the coupling constant 
in the weak coupling limit. Results for $c_m^{*}$ and $c_m^{\it cr}$ are 
compiled in Table \ref{discomega}.

 The discontinuity at the first order phase transition can be 
quantified in terms of the change in the trace of the Wilson line.
In weak coupling, $g^2 N_c/(4\pi) \ll 1$, we have
\begin{equation}
\label{tromegaexp}
 \tr \langle \Omega\rangle \ \approx  
  i \left({g^2 N_c \over 4 \pi}\right) {1 \over N_c} 
   \tr \left( \Omega_0 \vec{H} \cdot  \langle\vec{b}' \rangle \right),
\end{equation}
where $\Omega_0 = e^{i {2 \pi \over N_c} \vec{H} \cdot \vec{\rho}} $ is the  
center-symmetric holonomy.  The discontinuity $\Delta|\tr \langle 
\Omega \rangle|$ at $c_m^{\it cr}$, calculated from Eq.~(\ref{tromegaexp}) 
using the numerical value for $c_m^{\it cr}$, is also given in Table 
\ref{discomega}. The data for the discontinuity of the trace (with 
maximal trace normalized to unity) can be represented by the 
expression
\begin{equation}
\label{discomegafit}
\Delta \left| \frac{1}{N_c} \tr \Omega\right|  \simeq  
  \left(0.163  - \frac{0.085}{N_c} \right)\; {g^2 N_c \over 4 \pi} ~,
\end{equation}
which is valid in the large-$N_c$ limit with the parameters given
in Eq.~(\ref{parameters}) held fixed.

\begin{table}
\begin{center}
\begin{tabular}{c|c|c|c|c}
$N_c$ & $c_m^*=4 \sin^2 {\pi \over N_c}$ 
 & $c_m^{\it cr}$ &  $\frac{N_c^2}{4\pi^2}\times c_m^{\it cr}$
 & $\Delta |\tr \langle \Omega \rangle| \times \left({4 \pi \over g^2 N_c}
  \right)$ \\ \hline
3  & 3.000  & 2.45  & 0.56  & 0.37   \\
4  & 2.000  & 1.47  & 0.60  & 0.57 \\
5  & 1.382  & 0.97  & 0.61  & 0.75 \\
6  & 1.000  & 0.68  & 0.62  & 0.91 \\
7  & 0.753  & 0.50  & 0.62  & 1.07 \\
8  & 0.586  & 0.38  & 0.61  & 1.22 \\
9  & 0.467  & 0.30  & 0.61  & 1.37 \\
10 & 0.382  & 0.24  & 0.61  & 1.53 \\ 
\end{tabular}
\end{center}
\caption{\label{discomega}
Critical values $c_m^{*}$ and $c_m^{\it cr}$ for the endpoint of the 
metastable phase and the first order transition in $SU(N_c)$ gauge 
theory with one adjoint fermion on $\R^3\times \S_1$. We also show 
the critical coupling $c_m^{\it cr}$ scaled by $4\pi^2/N_c^2$, and 
give the discontinuity of the Wilson line in the weak coupling limit. }
\end{table}

\subsection{More on the (abelian)  large-$\mathbf{N_c}$ limit and (evading) 
the Hagedorn instability}
\label{sec:more}

 So far, we have  computed the critical mass numerically for $N_c=(3,4,
\ldots)$. The numerical results indicate that the critical value of the 
gaugino mass,  $m_{\it cr}$, is equal to  $m_{*}$ up to a factor of order  
one for arbitrary $N_c$.  This can be seen from the fourth column of 
Table \ref{discomega}, which shows the value of $c_m^{\it cr}$ scaled 
by the asymptotic value of $c_m^{*}$. Therefore, using Eq.~(\ref{m_cr_loc},
\ref{L_cr_loc}), we observe that $\frac{m^*}{\Lambda} \sim \frac{3}{4} 
(L \Lambda)^2$ at fixed $L$, or $L^* \sim  \Lambda^{-1} \sqrt {\frac{4m}
{3 \Lambda}}$ at fixed $m$. We conclude that in the  scaling regime the 
$SU(2)$ result derived in \cite{Poppitz:2012sw} has a smooth large-$N_c$ 
limit. 

\begin{figure}[t]
    {
    \parbox[c]{\textwidth}
        {
	\centering
	\begin{subfigure}{0.32\textwidth}
        \bc\includegraphics[angle=0, width=0.75\textwidth]{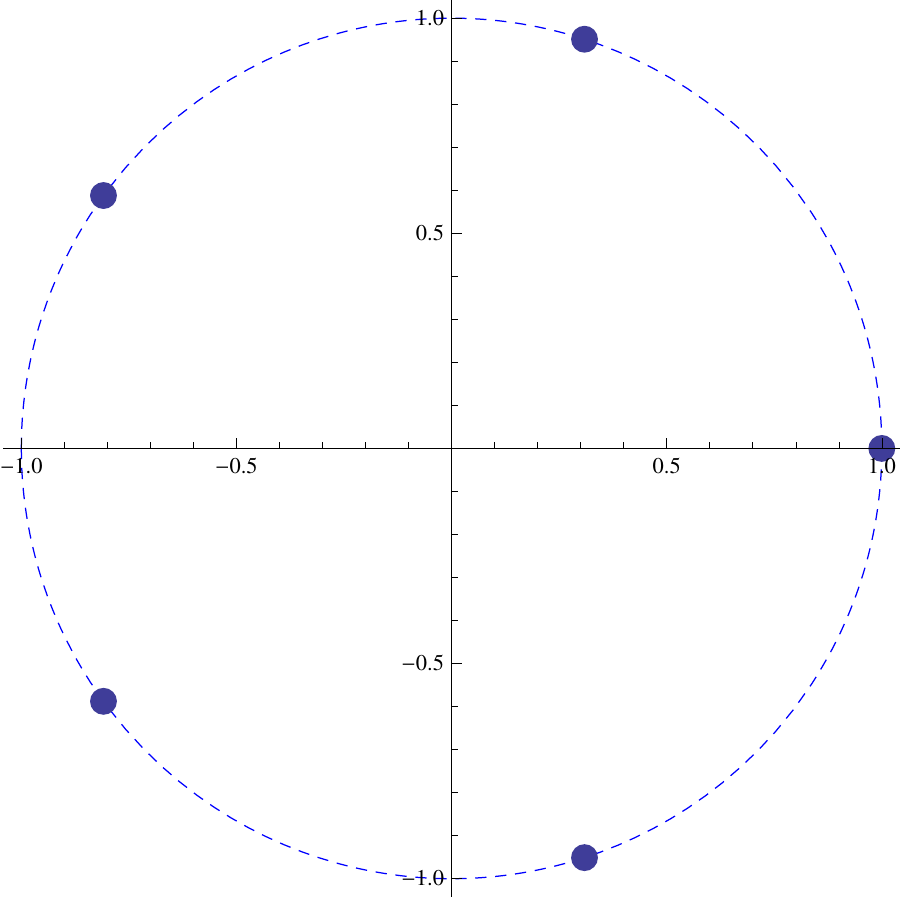}\ec
	\caption{$c_m \le c_m^{\it cr}$}
	\end{subfigure}
	\centering
	\begin{subfigure}{0.32\textwidth}
        \bc\includegraphics[angle=0, width=0.75\textwidth]{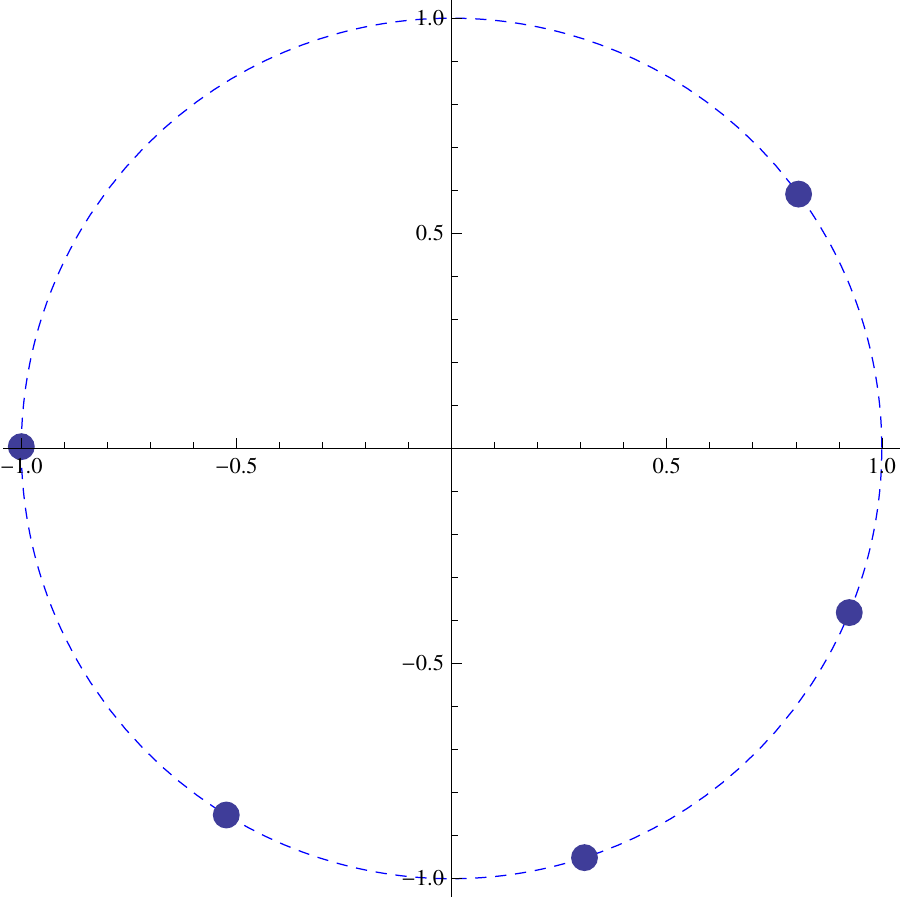}\ec
	\caption{$c_m = c_m^{\it cr} \approx 0.97$ }
	\end{subfigure}
	\centering
	\begin{subfigure}{0.32\textwidth}
        \bc\includegraphics[angle=0, width=0.75\textwidth]{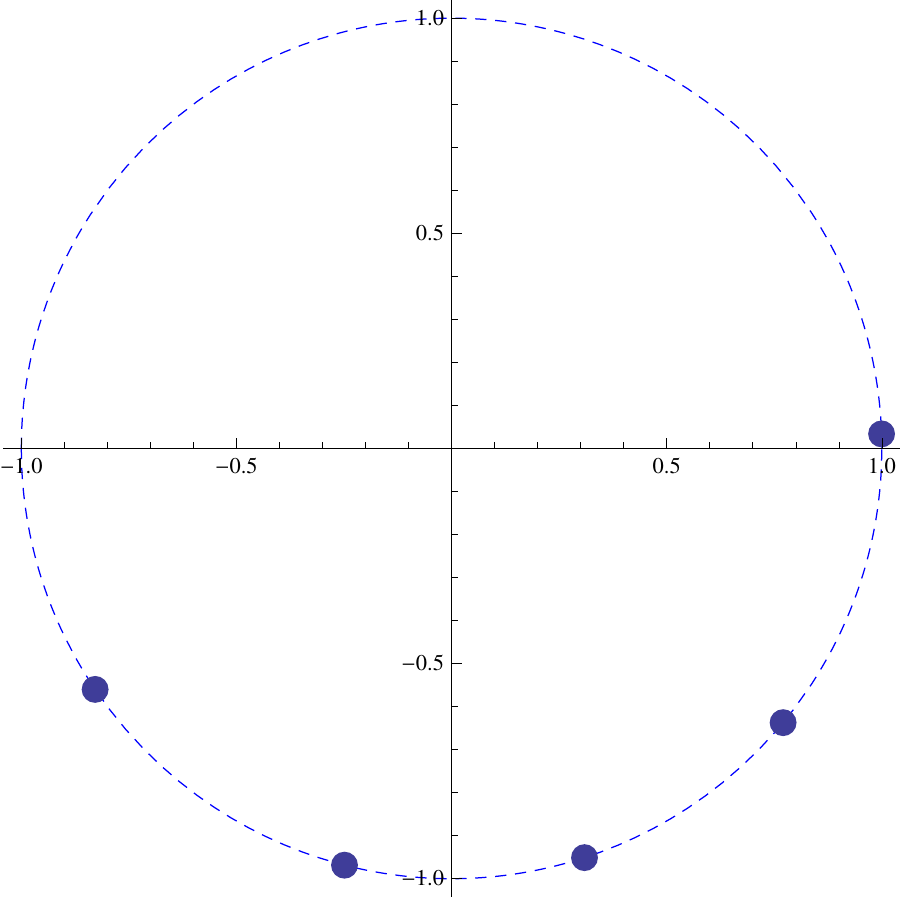}\ec
	\caption{$c_m = 2 c_m^{\it cr}$  }
	\end{subfigure}
	\hfil
	\caption{\small 
        The distribution of the eigenvalues of the Polyakov loop around 
        $\S^1_L$ for $N_c = 5$ for different values of $c_m$, shown for 
        $g^2 N_c = 0.1$.}
        \label{fig:su5}
        }
      }
\end{figure}

\begin{figure}[t]
    {
    \parbox[c]{\textwidth}
        {
	\centering
	\begin{subfigure}{0.32\textwidth}
        \bc\includegraphics[angle=0, width=0.75\textwidth]{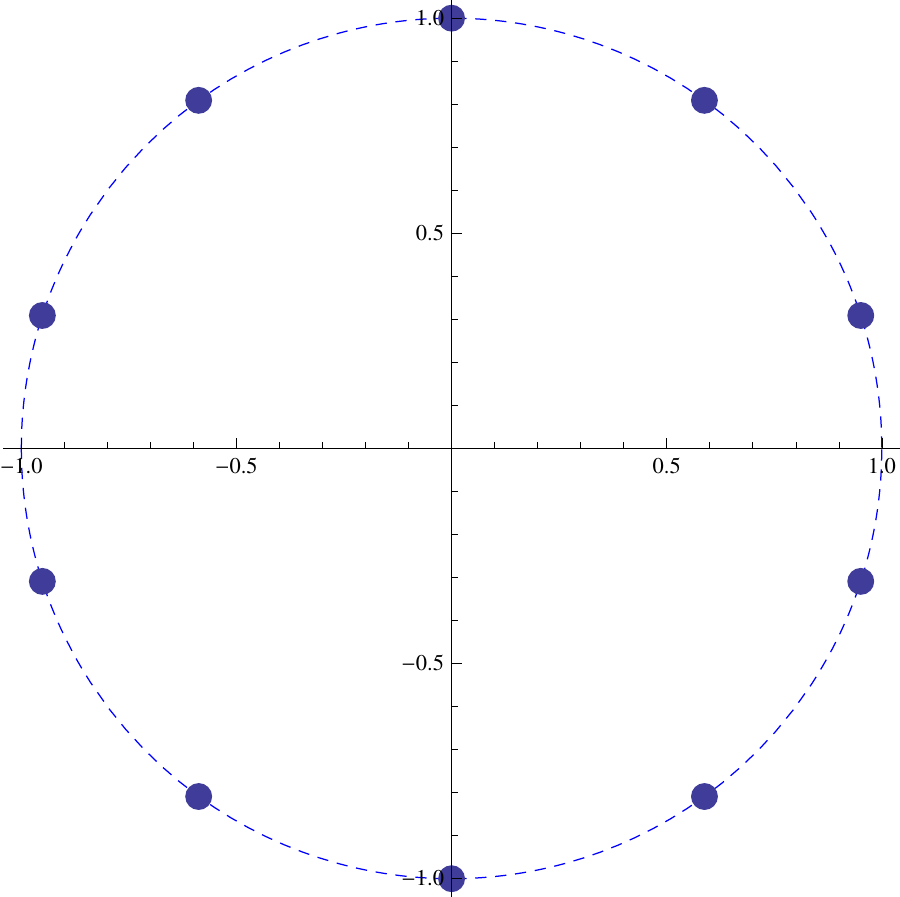}\ec
	\caption{$c_m \le c_m^{\it cr}$}
	\end{subfigure}
	\centering
	\begin{subfigure}{0.32\textwidth}
        \bc\includegraphics[angle=0, width=0.75\textwidth]{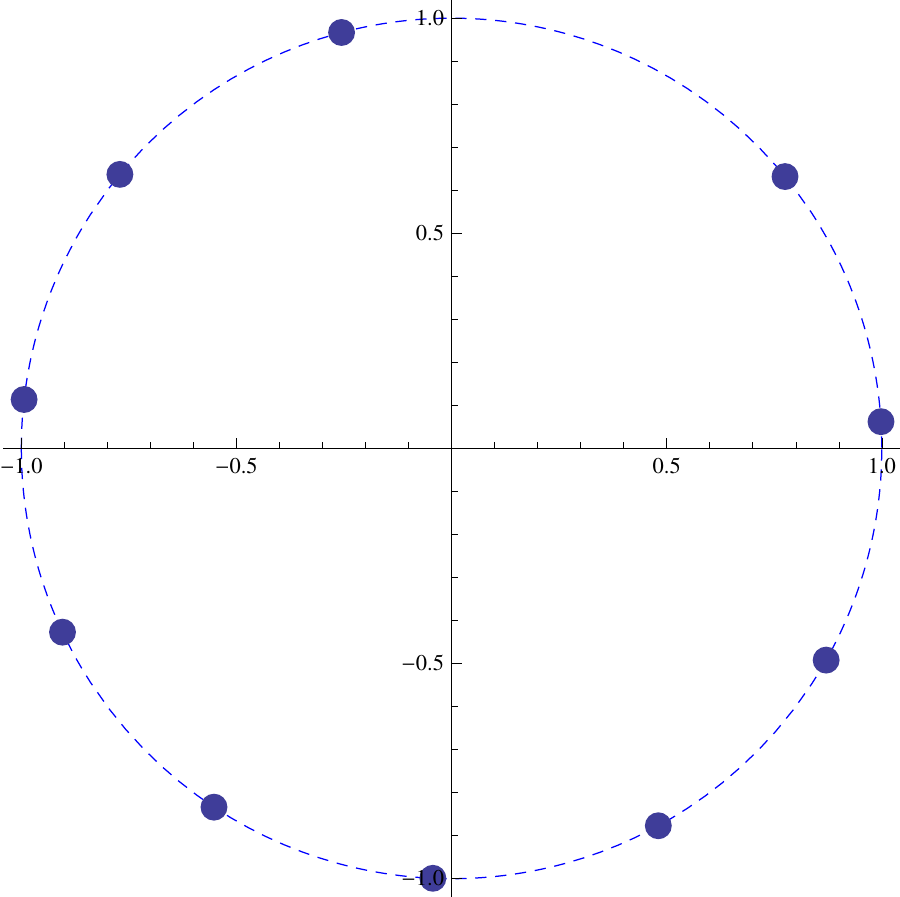}\ec
	\caption{$c_m = c_m^{\it cr} \approx 0.243$ }
	\end{subfigure}
	\centering
	\begin{subfigure}{0.32\textwidth}
        \bc\includegraphics[angle=0, width=0.75\textwidth]{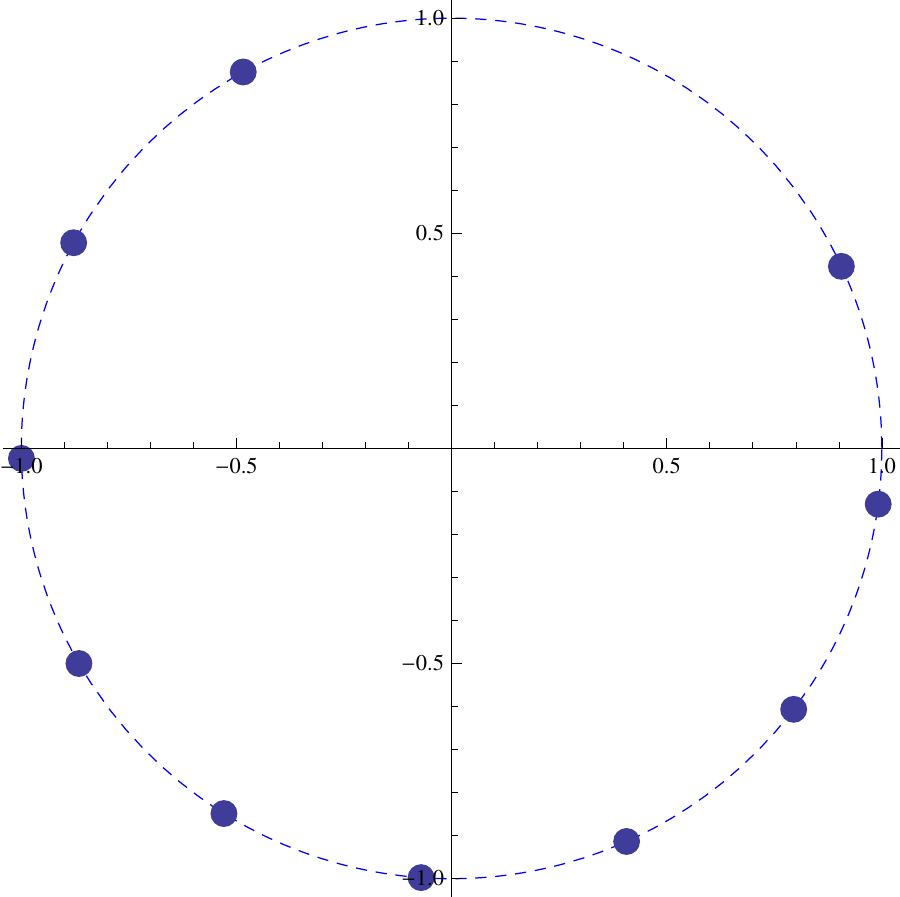}\ec
	\caption{$c_m = 2 c_m^{\it cr}$  }
	\end{subfigure}
	\hfil
        \caption{\small The distribution of the eigenvalues of the Polyakov 
        loop around $\S^1_L$ for $N_c = 10$ for different values of $c_m$, 
        shown for $g^2 N_c = 0.1$.}
        \label{fig:su10}
        }
        }
\end{figure}

 We note that in the abelian confinement regime the size of the dual 
circle $\widetilde \S^1$ on which the eigenvalues reside is equal to 
$1/L$, which grows linearly with $N_c$, while the separation of the 
eigenvalues remains fixed, $2\pi/(LN_c) \sim O(N_c^0)$. This is in 
sharp contrast with the non-abelian confinement regime and the ordinary 
large-$N_c$ limit. In the latter, $1/L=O(N_c^0)$ and the separation 
between eigenvalues is $2 \pi/(LN_c) \sim O(N_c^{-1})$, forming a dense 
set in perturbation theory. In the latter case, since $2\pi/(LN_c) 
\ll \Lambda$, all the low momentum modes are strongly coupled, and 
the eigenvalues are uniformly distributed over the unit circle.  At 
the critical point in the weakly coupled regime, the uniform separation 
between eigenvalues exhibits a jump into a non-uniform one, by opening 
a ``gap'' on top of the usual one. As one increases $c_m$ the gap 
continues to grow, as shown in Figures~\ref{fig:su4},\ref{fig:su5}, 
and \ref{fig:su10} for $SU(4),SU(5)$, and $SU(10)$, respectively. In 
particular, we did not observe, in the semi-classical regime, an 
instability towards partial center-symmetry breaking phases, similar 
to those found in deformed Yang-Mills theory and massive QCD(adj),  
\cite{Myers:2007vc, Unsal:2010qh}, see Ref.~\cite{Ogilvie:2012is} 
for a review.

There is one more interesting issue that appears in the large-$N_c$ limit.  
The density of states of large-$N_c$ gauge theories is expected to 
exhibit an exponential growth, $\rho(E) \sim  e^{\beta^*  E} = e^{ E/T_H}$, 
where $\beta^*\equiv T_H$ is the Hagedorn temperature. This idea is 
related to the conjecture that large-$N_c$ gauge theory is dual to a weakly 
coupled string theory whose density of states is known to grow exponentially.
Then, $Z(\beta) = \int^{\infty} dE \; e^{(\beta^* -\beta) E}$ diverges for $\beta 
< \beta^*$, indicating the existence of a limiting temperature for the 
confined phase. As discussed above $T_H$ is expected to be larger than
$T_c$. In simulations of $SU(12)$ Yang-Mills it was found that $T_H= 
1.116T_c$ \cite{Bringoltz:2005xx}.  This seems to imply that it is  
impossible to extend the confined phase to the regime of small-$L$ and
weak coupling. Naively, even if one evades deconfinement, it seems unlikely  
that one can  evade  the Hagedorn instability.  It is noteworthy to 
understand how our construction avoids this obstacle. 
 
 The bosonic and fermionic density of states $\rho_{\cal B}(E)$ and 
$\rho_{\cal F} (E)$ are properties of the Hamiltonian, independent 
of the boundary condition along the $\S_1$. The crucial point is that 
 the density operator  $e^{- \beta H}$ of the thermal compactification 
is replaced by a $\Z_2$-graded density operator, $e^{- \beta H} (-1)^F $, 
whose matrix elements, unlike the ordinary density matrix, are not 
necessarily positive definite.  The  twisted partition function is 
of the form $ \widetilde Z (L) = \int^{\infty} dE \; ( \rho_{\cal B}(E)  
-  \rho_{\cal F} (E) ) e^{- L E}$, so that even if the density of states
grows exponentially the spatially compactified theory 
need not have a limiting scale. 
This is most transparent in the supersymmetric limit, where all positive  
energy states are paired bosonic and fermionic states, and  $\widetilde Z 
(L)$ is the supersymmetric index.\footnote{
These considerations are not restricted to supersymmetry theories, 
or theories with softly broken supersymmetry. For example, $N_c=\infty$ 
non-supersymmetric QCD with multiple adjoint fermions does not possess 
{\it any} phase transition upon circle compactification, and satisfies 
large-$N_c$ volume independence, see \cite{Unsal:2010qh,Lucini:2012gg} 
and references therein.} 
In the softly broken supersymmetric theory, we find a phase transition  
at the scale given in Eq.~(\ref{L_cr_loc}), $L^* \sim  \Lambda^{-1} \sqrt{m\over \Lambda}\ll {2 \pi \over \Lambda N_c}$ (recall (\ref{parameters})), 
parametrically smaller than the Hagedorn scale $\beta^{*} \sim \Lambda^{-1} $.   
Strictly speaking, this  is a quantum phase transition, which  should not 
be interpreted thermally for small $m$.  However, it is continuously 
connected to the thermal deconfinement transition, and this makes the 
semi-classical analysis feasible. 

\section{$\theta$-dependence of the phase transition and topological 
interference}
\label{sec:theta}

  In the Minkowski-space formulation of the path integral, the $\theta$-term 
generates an extra phase of the path amplitude, $e^{i S(\gamma)} \rightarrow 
e^{i S(\gamma) + i \theta Q(\gamma)}$, where $\gamma$ is the path connecting the 
initial and final field configurations. This is analogous to the Aharonov-Bohm 
(AB) phase in quantum mechanics, and the $\theta$-angle can be viewed
as an AB-flux. Turning on the $\theta$ angle generates interference among 
path amplitudes which affects physical observables and the vacuum structure 
of the theory. 
   
 In the case of Yang Mills theory analytic continuation of the path integral 
to Euclidean space induces a positive definite action for $\theta=0$. However, 
for $\theta \neq 0$, the theory has a sign problem, preventing straightforward 
numerical simulations, see Ref.~\cite{D'Elia:2012vv,Vicari:2008jw} and 
references therein. In the semi-classical regime the sign problem manifests 
itself as a complex fugacity for topological configurations. For example, 
the 4d instanton fugacity $e^{-S_I}$ is modified to $e^{-S_I + i \theta}$. The 
density of topological configurations remain unaltered, but they acquire 
a pure phase depending on the topological charge. This leads to qualitative 
and quantitative changes in the vacuum structure of the theory, which we 
refer to as {\it topological interference} effects. In particular, the 
fugacity of 1-defects (monopole-instantons) becomes  
complex, but the fugacity of topologically neutral bions remains 
unchanged. In the case of magnetic bions the result depends on the 
gauge group $G$ --- for $SU(N_c)$ the fugacity remains unaltered, but 
for $G_2$ it acquires a phase depending on the topological charge of 
the magnetic bion. 

There is a subtle point here (for an earlier related discussion, see \cite{Thomas:2011ee}). The action of the monopole-instanton is 
modified as $\exp(-S_0)\rightarrow \exp(-S_0 + \frac{i \theta}{N_c})$
with $S_0=8\pi^2/(g^2N_c)$. Since $\theta$ is a periodic variable with 
period $2\pi$, this implies that there exists a family of Lagrangians:
\begin{align}
\label{vacuum-family}
{\cal L}^{(k)} = {\cal L}_{kin.} + V_{np}^{(k)}, \qquad k=0, \ldots, N_c-1  
\end{align}
where the $N_c$ potentials $V_{np}^{(k)}$ are given in Eq.~(\ref{V_np_Nc}), 
and the kinetic term ${\cal L}_{kin.}$ in Eq.~(\ref{kinetic2}). For a given 
value of $\theta$, we must determine the correct effective long-distance 
theory. The correct theory  is determined by  comparing the vacuum 
energy density associated with each branch, and selecting the branch
with the minimum vacuum energy. The theta dependence of the vacuum energy 
density, at leading order in semi-classical expansion, is given by 
\begin{align}
\label{vacuum}
 { \cal E} (\theta) 
     = \min_{k}  { \cal E}_k (\theta)  
     \equiv \min_{k} \left[ - \frac{V_{mon}^0}{L}  
          \cos\left({2 \pi k + \theta \over N_c}\right) \right]\, , 
\end{align}
which arises from monopole-instantons. The contribution due to neutral and 
magnetic bions cancel each other exactly, and the analytic contribution 
of 4d instantons is exponentially suppressed. Eq.~(\ref{vacuum}) implies 
that the vacuum energy is analytic everywhere except for odd integer 
multiples of $\pi$ where it exhibits non-analyticity.  For example, in 
the range $\theta \in (-\pi, \pi)$ the ground state energy density is 
${\cal E}_0 (\theta)$, while for $\theta \in (\pi, 3\pi)$ the vacuum energy 
density is ${\cal E}_1 (\theta)$. This implies that at $\theta=\pi$ the 
$k=0$ and $k=1$ branches become degenerate. This degeneracy that we see 
in the long-distance effective theory is a manifestation of the  CP-symmetry 
in the microscopic theory at $\theta=\pi$. Thus, the vacuum structure and 
the symmetry realization for general group  $G$ is modified at $\theta=\pi$ 
as
\begin{align} 
\theta=\pi:  \qquad   Z(G) \times Z_2^{\rm CP}  \Longrightarrow  
   \left\{\begin{array}{ll} 
       Z(G) \times 1       & {\rm confined} \\ 
       1 \times Z_2^{\rm CP}   & {\rm deconfined}
\end{array} \right.
\end{align} 
For all gauge groups, due to spontaneous breaking of CP, the vacuum 
in the confined phase is two-fold degenerate at $\theta=\pi$. Note that $Z_2^{\rm CP}$ 
also gives a symmetry that can be used to distinguish the phases of $G_2, 
F_4,E_8$ at  $\theta=\pi$. At very high $T$, CP is restored and at low $T$
it is broken. 

Using Eq.~(\ref{vacuum-family}), we can investigate the $\theta$ dependence 
for general observables, in particular the dependence of the phase transition 
and the critical scale on $\theta$. It is not hard to see that the theta 
dependence is most pronounced for gauge groups of small rank. We first 
consider the case of $SU(2)$. We will show that  $L^*(\theta)$ assumes 
its maximum at $\theta=\pi$, where the monopole-instantons events interfere 
destructively, resulting in a larger $L^{\it cr}$ for breaking of the center 
symmetry. In the limit where the rank is large, $T_c$ becomes theta
independent. 

\subsection{$\mathbf\theta$-dependence  of $\mathbf{c_m^{\it cr}}$ for 
$\mathbf{N_c=2}$}

 In the case $N_c=2$ the $\theta$-dependence of $c_m^{\it cr}$ can be 
studied quite explicitly (the $N_c>2$ case was recently studied in \cite{Anber:2013sga}). We take the $SU(2)$ roots as $\alpha_1 = 
- \alpha_2 = \sqrt{2}$ and find, from the potential given in 
Eq.~(\ref{totalv}):
\begin{eqnarray}
\label{su2potentialtheta}
{V(b', \sigma')\over V_{bion}^0} 
  &=& 2 \cosh\left( 2 \sqrt{2} b'\right) 
       - 2 \cos\left( 2 \sqrt{2} \sigma'\right) \nonumber \\
  & & - c_m \left(1 + {g^2 \over 4 \pi^2}\sqrt{2} b'\right) e^{- \sqrt{2} b'} 
      \cos\left( \sqrt{2} \sigma' + {\theta \over 2} +  \pi k\right) 
             \nonumber \\
  & & - c_m\left(1 - {g^2 \over 4 \pi^2}\sqrt{2} b'\right) e^{ \sqrt{2} b'} 
      \cos\left( \sqrt{2} \sigma' - {\theta\over 2} -  \pi k\right)~,
\end{eqnarray}
where the various factors of $\sqrt{2}$ are due to the periodicities 
given in Eq.~(\ref{periodicities}) and the normalization of the roots.
Note that $k=0,1$ labels the choice of supersymmetric vacuum, corresponding 
to $\Z_{4} \rightarrow \Z_2$ $R$-symmetry breaking. There is no $R$-symmetry 
when $c_m$ is non-zero, and  the vacuum branch is the $k=0$ one when $\theta 
\in (-\pi, \pi)$. For the angular range $\theta 
\in (\pi, 3\pi)$, the vacuum branch is $k=1$.
Thus, proceeding with $k=0$ and considering subcritical $c_m$, we find 
that for $\theta < \pi$ the global minimum of Eq.~(\ref{su2potentialtheta}) 
is at $\langle \sigma \rangle = 0$, $\langle b^\prime \rangle = 0$. For $\theta 
> \pi$ it is located at $\langle \sigma \rangle = {\pi\over \sqrt{2}}$, 
$\langle b^\prime \rangle = 0$, while for $\theta=\pi$ the two minima are 
degenerate, and the theory spontaneously violates CP.

 Consider the range $0 \le \theta \le \pi$ (similar arguments apply to
other values of $\theta$). Expanding the potential to quadratic order
around the minimum $\langle \sigma \rangle = 0$, $\langle b^\prime \rangle 
= 0$ gives
\begin{equation}
\label{quadraticV}
8 (\delta b \; \delta\sigma) \left[ \left(
   \begin{array}{cc} 1 & 0 \cr 0 & 1 \end{array} \right) 
- {c_m\over 4} \left(
   \begin{array}{cc}
      \left(1 - {g^2\over 2 \pi^2}\right)\cos{\theta\over 2}  
   &  \left(1 - {g^2\over 4 \pi^2}\right)\sin{\theta\over 2} \cr  
      \left(1 - {g^2\over 4 \pi^2}\right)\sin{\theta\over 2}  
   & -\cos{\theta\over 2} 
   \end{array} \right)\right]  
   \left( \begin{array}{c} \delta b \cr \delta\sigma \end{array}\right)~.
\end{equation}
One of the two eigenvalues of the matrix inside the square brackets in 
Eq.~(\ref{quadraticV}) can change sign:
\begin{equation}
\label{evs}
   1 \pm {c_{m} \over 4 } 
  \left(1 - {g^2 \over 4 \pi^2}\left( 1\mp \cos {\theta\over 2}\right)\right) ~,
\end{equation}    
where the  signs are correlated. Thus, a negative eigenvalue appears at 
the critical value $c_m^*$. Extending the angular range to arbitrary $\theta$, 
\begin{equation}
  c_m^{\it *}   
   =  \max_{k=0,1} \left[ 4 + {g^2 \over \pi^2}
 \left(1 + \cos \left( {\theta\over 2 } + \pi k  \right) \right) 
 + \ldots~ \right],
\label{criticalltheta}
\end{equation}
where ellipsis represent terms with additional  $g^2$-suppression. Note  
that within the calculable regime, the critical $c_m$ given in 
Eq.~(\ref{criticalltheta}) only acquires $\theta$ dependence due to the 
$g^2$-suppressed terms.  This is special to the $N_c=2$ case, for 
$N_c \geq 3$, $\theta$ dependence appear at the leading order.

Using (\ref{criticalltheta}), we obtain, like vacuum energy density or 
any other non-perturbative observable, a multi-branched function, which 
is  two-branched for $SU(2)$, given by
\begin{equation}
  L^{\it *} (\theta)   =    \min_{k=0,1}  { L}_k^* (\theta)   = 
  L^{\it *} (0) 
    \min_{k=0,1} \left[ 4 + {g^2 \over \pi^2}\left(1 + \cos \left( {\theta\over 2 } + \pi k  \right) \right) + \ldots~ \right]^{-{1 \over 2}}, 
\label{criticalltheta2}
\end{equation}
Both $L_0^*(\theta)$ and  $L_1^*(\theta)$ are  $4\pi$ periodic, whereas 
the  $\min_{k=0,1}  { L}_k^{*} (\theta)$ is $2\pi$ periodic as it must. 
 
The physical interpretation of the two branches is as follows.  For a  
given value of $\theta$ the potential in Eq.~(\ref{su2potentialtheta})
always has two extrema. Both of these extrema are center-symmetric  
minima for small $g$ and are located  at $\langle \sigma \rangle = 0, 
\; {\pi\over \sqrt{2}}$.  For $\theta \neq \pi$, one of these minima is 
the true vacuum, say $k=0$  and the other is meta-stable,  say $k=1$. 
$L_0^*(\theta)$ and   $L_1^*(\theta)$ are the respective transition 
scales obtained by studying the theory on true vacuum  versus  meta-stable 
vacuum. At $\theta=\pi$, the two vacua become degenerate, leading to a 
cusp (non-analyticity) in  $ L^{\it *} (\theta)$, where $L_0^*(\theta)=
L_1^*(\theta)$.

 As Eq.~(\ref{criticalltheta}) shows, as $\theta$ increases away from zero 
to $\pi$, the critical  $c_m^{\it cr}$ decreases. At fixed $m$ decreasing 
$c_m^{\it cr}$ implies increasing $L^{\it cr}$ in the range $\theta \in [0,\pi)$.
As in \cite{Unsal:2012zj}, the increase of $L^{\it cr}$ with increasing 
$\theta$ can be attributed to ``topological interference". In our 
context, the physics of topological interference is that the effect of 
monopole-instantons is suppressed  at $\theta=\pi$ with respect to $\theta=0$ 
by destructive interference  among path histories. As in Aharonov-Bohm phase, 
this extra topological phase factor gives additional interference among path 
histories (either Euclidean or Minkowski). Therefore, at $\theta=\pi$, the 
center-destabilizing potential due to monopole-instantons is suppressed 
with respect to  $\theta=0$. Comparing $ \theta=\pi $ to  $\theta=0$ it 
requires stronger coupling, meaning a larger value of $L$, in order for
the monopole-instantons to overcome the center-stabilizing 
($\theta$-independent) effect of the neutral bions.  

 If the  weak coupling behavior holds qualitatively at strong coupling 
(i.e., for  decoupling values of the gaugino mass), an increasing 
$L^{\it cr}(\theta)$ ($\beta^{\it cr}(\theta)$) implies that the deconfinement 
temperature $T_c (\theta)$ decreases with increasing  $\theta \in [0, \pi]$.  
This behavior was discussed in \cite{Unsal:2012zj} and recently observed 
in lattice simulations \cite{D'Elia:2012vv}.

\section{$\mathbf{G_2}$: First order transition without symmetry breaking} 
\label{sec:g2}

 $G_2$ Yang Mills theory is an interesting test case for 
studying the deconfinement transition \cite{Holland:2003jy,Pepe:2006er,Greensite:2006sm,Cossu:2007dk,Diakonov:2010qg,Ilgenfritz:2012wg,Dumitru:2012fw}.
$G_2$, along with $F_4$ and $E_8$, has no global (center) symmetry and hence no 
obvious order parameter for deconfinement.  
Lattice calculations show
that there is a first order phase transition between a low temperature
``confined'' phase, and a high temperature ``deconfined'' phase. In the 
confined phase there is an effective string potential at intermediate 
distances, but at very large distance the potential between fundamental
charges is screened. The Wilson line is small but non-zero in the 
confined phase, and changes discontinuously at the phase transition. 
In the high temperature phase the Wilson line is close to one.\footnote{As 
discussed above the theory has a global  $\Z_2$ (CP)symmetry at 
$\theta=\pi$. This symmetry is broken in the confined phase. It is easy 
to show that it is unbroken in the very high-$T $ deconfined phase. 
However, it is not a priori obvious that CP restoration and the 
discontinuity of the Wilson line occur at the same critical temperature. }

\begin{figure}[t]
    {
    \parbox[c]{\textwidth}
        {
	\centering
	\begin{subfigure}{0.32\textwidth}
        \bc\includegraphics[angle=0, width=0.75\textwidth]{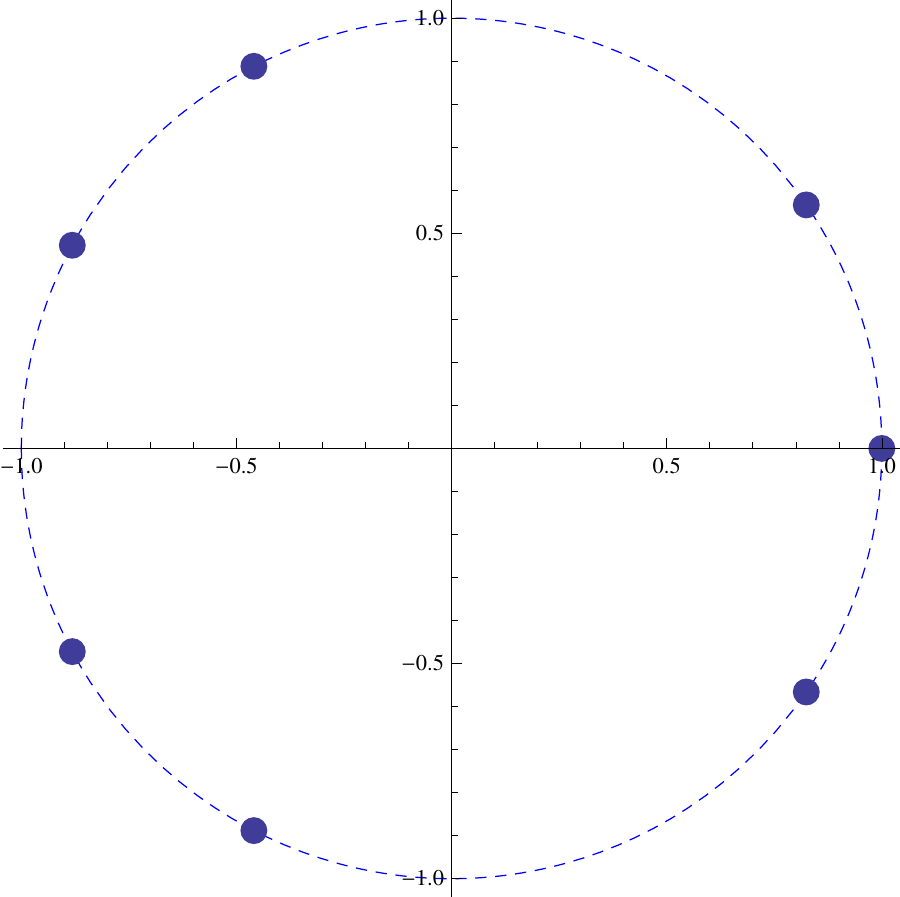}\ec
	\caption{$c_m \le c_m^{\it cr}$}
	\end{subfigure}
	\centering
	\begin{subfigure}{0.32\textwidth}
        \bc\includegraphics[angle=0, width=0.75\textwidth]{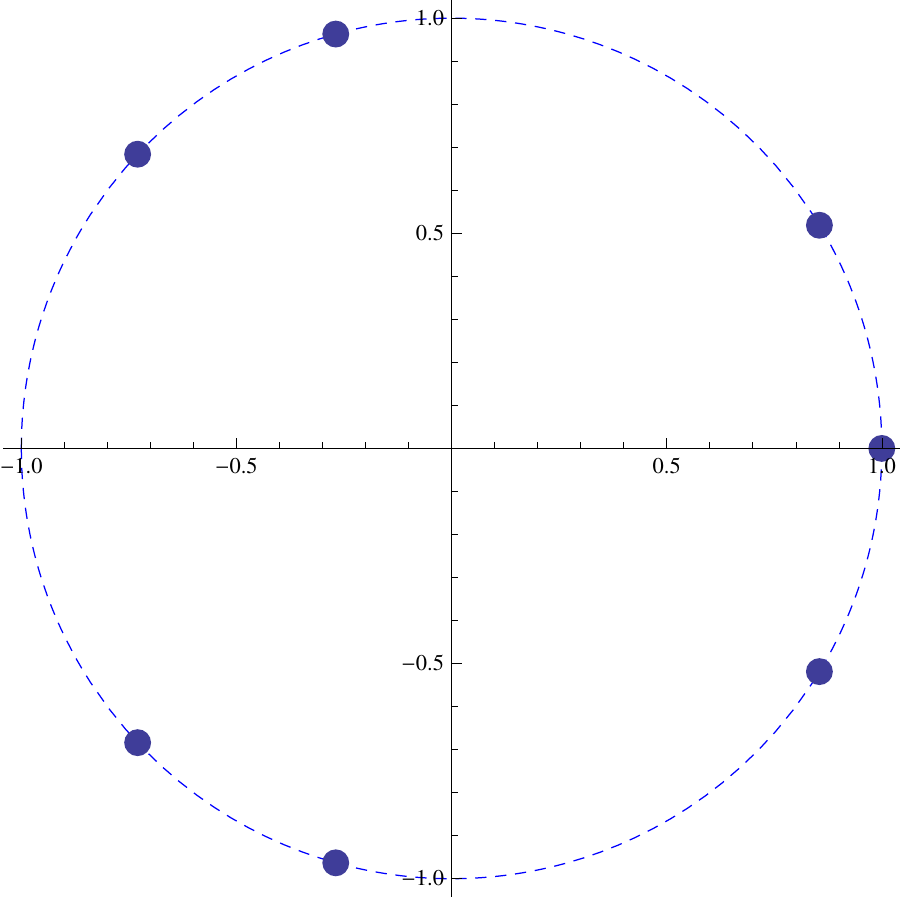}\ec
	\caption{$c_m = c_m^{\it cr} \approx 1.587$ }
	\end{subfigure}
	\centering
	\begin{subfigure}{0.32\textwidth}
        \bc\includegraphics[angle=0, width=0.75\textwidth]{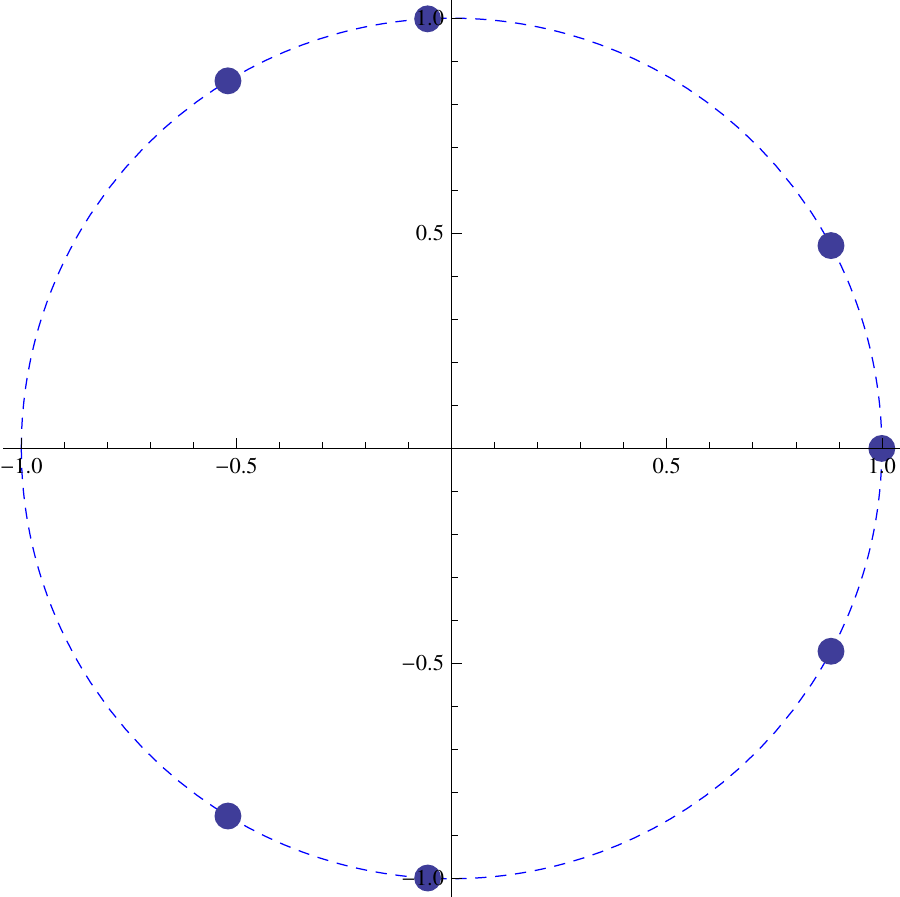}\ec
	\caption{$c_m = 2 c_m^{\it cr}$  }
	\end{subfigure}
	\hfil
	\caption{\small 
        The distribution of the eigenvalues of the Polyakov loop around 
        $\S^1_L$ for a $G_2$ gauge theory, for subcritical, critical, and 
        supercritical values of $c_m$ (plotted with ${g^2\over 4 \pi }= 
        0.2$ in order to visualize the jump).}
        \label{fig:g2eigen}
        }
      }
\end{figure}

 The group $G_2$ has rank two. The fundamental representation is 
seven dimensional, and the adjoint representation has dimension 14.
The tensor product of three 14 dimensional representations contains
the fundamental representation, $[14]\times[14]\times[14]=[1]+[7]+
\ldots$, which explains why fundamental charges are screened. As in 
$SU(N_c)$ SYM theory on $\R^3 \times \S^1$ dynamical abelianization, 
$G_2 \rightarrow U(1)^2$, is  a consequence of the existence of neutral 
bions, which generate a repulsion among the eigenvalues of Wilson line, 
see Fig.~\ref{fig:g2eigen}.\footnote{In phenomenological models it is 
straightforward to generate eigenvalue repulsion in $SU(N_c)$, but it 
is not easy to find a non-perturbative potential that leads to eigenvalue
repulsion in $G_2$, see Section VIII.C in~\cite{Dumitru:2012fw}. We 
find that the neutral bion potential automatically gives eigenvalue
repulsion for all gauge groups.}
The long distance theory contains the two dual photons $\vec{\sigma}$, 
their scalar superpartners $\vec\phi$ (the massless components of the 
gauge field along the $\S^1$), and fermionic superpartners.\footnote{
The dual simple roots $\vec\alpha_1^*$, $\vec\alpha_2^*$, and the affine 
root $\vec\alpha_0^* = - 2\vec\alpha_1^* - \vec\alpha_2^*$ (two dimensional 
vectors) have length squared $2$ ($\vec\alpha_1^*$), $6$ ($\vec{\alpha_2}^*$), 
and $2$ ($\vec\alpha_0^*$),  and obey $\vec\alpha_1^* \cdot \vec\alpha_2^* 
= -3$, $\vec\alpha_1^* \cdot \vec\alpha_0^*=-1$, $\vec\alpha_2^* \cdot 
\vec\alpha_0^* = 0$. The fundamental weights $\vec\omega_j$  obey $\vec
\omega_i \cdot \vec\alpha_j^* = \delta_{ij}$ ($i,j=1,2$). The Weyl vector 
$\vec\rho = \vec\omega_1 + \vec\omega_2$ satisfies $\vec\rho \cdot \vec
\alpha_{1,2}^* = 1$. The fundamental representation Cartan generators are 
normalized as $\tr H_a H_b = 2 \delta_{ab}$ and are explicitly given by 
$H_1 = {1\over \sqrt{2}} {\rm diag} (1, -1, 0, -1, 1, 0, 0)$, $H_2 = 
{1 \over \sqrt{6}} {\rm diag} (1, 1, -2, -1, -1, 2, 0)$.}
The fundamental Wilson line along $\S^1$ is
\begin{equation}
\label{omegag2}
\Omega = e^{ i \vec{H} \cdot \vec\phi}~.
\end{equation}
Similar to the $SU(N_c)$ case, we define the fluctuations $\vec{b}'$ and 
$\vec\sigma'$ of the fields $\vec\phi$ and $\vec\sigma$ around the $k$-th 
supersymmetric ground state
\begin{eqnarray}
\label{fieldsg2}
\vec{\phi} &=& \left({\pi \over 2} 
    + {g^2 \over 16 \pi} \log {4 \over 3} \right) \vec\rho 
    - {g^2 \over 4 \pi} \left(\vec\omega_1 \log 2 -\vec\omega_2 \log 3 \right)
    + {g^2 \over 4 \pi} \; \vec{b}'  \nonumber \\
\vec\sigma + {\theta \vec\phi \over 2 \pi} 
      &=& {\theta + 2 \pi k \over 4} \; \vec\rho + \vec\sigma'~.
\end{eqnarray}
Here $k=1,2,3,4$ corresponds to a choice of vacuum breaking the discrete 
$R$ symmetry of the supersymmetric theory, $\Z_8 \rightarrow \Z_2$.  One 
difference from the $SU(N_c)$ case is that the expectation value of the 
fundamental Wilson loop Eq.~(\ref{omegag2}) in the supersymmetric ground  
state $\langle\vec\sigma' \rangle = \langle\vec{b}\rangle = 0$ is nonzero:
\begin{eqnarray}
\label{susypolyakovvev}
\tr \langle \Omega\rangle &=& \tr\; 
    e^{\; i {\pi \over 2}  \vec{H}\cdot \vec\rho+ i {g^2 \over 4 \pi} \vec{H} 
        \cdot\left({1 \over 4 }\vec\rho \log{4\over 3} 
         - \vec\omega_1 \log 2 + \vec\omega_2 \log 3\right) } \nonumber\\
&\approx & \tr \; e^{\; i {\pi \over 2}  \vec{H}\cdot \vec\rho} 
        \left(1 + i \;{g^2 \over 4 \pi} \vec{H} \cdot 
         \left({1 \over 4 }\vec\rho \log{4\over 3} 
         - \vec\omega_1 \log 2 + \vec\omega_2 \log 3\right) \right)  \\
&=& {g^2 \over 4 \pi} \left({\pi \log 2 \over \sqrt{3}}  
  - {\pi \log 3\over 2}  \right) \approx -0.15\; {g^2 \over 4 \pi}~.\nonumber
\end{eqnarray}
This should not come as a surprise, as there is no center symmetry  
requiring  $\tr \langle\Omega\rangle=0$ in  the confining phase. 

The nonperturbative potential for the $\vec{b}'$ and $\vec\sigma'$ is 
due to neutral bions, magnetic bions, and monopole-instantons. The 
result can be derived using the methods introduced in \cite{Poppitz:2012sw},
based on the non-perturbative superpotential given in \cite{Davies:2000nw}
and the field redefinition in Eq.~(\ref{fieldsg2}). We find
\begin{eqnarray}
\label{g2potential}
 {V_{total} \over V_{bion}^0}
  &=&  4 e^{- 2 \vec{b}^\prime \cdot \vec\alpha_1^*} 
    + 3  e^{- 2 \vec{b}^\prime \cdot \vec\alpha_2^*} 
    +  e^{- 2 \vec{b}^\prime \cdot \vec\alpha_0^*}  \nonumber \\
&&  -  6 e^{-  \vec{b}^\prime \cdot ( \vec\alpha_1^* + \vec\alpha_2^*) } 
         \cos(\vec\alpha_1^* - \vec\alpha_2^*)\cdot \vec\sigma^\prime   
    -  2 e^{ - \vec{b}^\prime \cdot ( \vec\alpha_0^* + \vec\alpha_1^*) }  
           \cos(\vec\alpha_0^* - \vec\alpha_2^*) 
           \cdot \vec\sigma^\prime \nonumber \\[0.1cm]
&& - c_m \left\{ \left( 2 
     +  {g^2 \over 2 \pi^2}  \left({2 \vec{b}^\prime \cdot \vec\alpha_1^* 
                 - {1\over 2} \log 12}\right)\right)\; 
     e^{- \vec{b}^\prime \cdot \vec\alpha_1^*} 
     \cos\left( \vec\sigma^\prime \cdot \vec\alpha_1^* 
               + {\theta + 2 \pi k \over 4} \right)  \right. \\
&&\left.   \qquad + \left( 1 
     +{g^2\over 2 \pi^2}\left(  \vec{b}^\prime \cdot \vec\alpha_2^* 
                 + {1\over 4} \log 108 \right) \right) \;
     e^{- \vec{b}^\prime \cdot \vec\alpha_2^*} 
        \cos\left( \vec\sigma^\prime \cdot \vec\alpha_2^* 
               + {\theta + 2 \pi k \over 4} \right) \right. \nonumber \\ 
&&\left.  \qquad + \left( 1 
     + {g^2 \over 2 \pi^2} \left(  \vec{b}^\prime \cdot \vec\alpha_0^* 
             - {1\over 4} \log {3 \over4 } \right) \right) \; 
     e^{- \vec{b}^\prime \cdot \vec\alpha_0^*} 
        \cos\left( \vec\sigma^\prime \cdot \vec\alpha_0^* 
               + {\theta + 2 \pi k \over 4} \right) 
\right\}~. \nonumber
\end{eqnarray}
Similar to the $SU(N_c)$ case, we use $V_{bion}^0$ to denote the overall 
strength of the bion potential, while $c_m$ is the ratio of the strengths 
of monopole to bion potentials. These parameters are given by
\begin{equation}
\label{g2potentials}
V_{bion}^0 = {16 \pi^2 \sqrt{3} \over g^2} L^3 \Lambda^6~, ~~ 
 c_m =  {m \over L^2 \Lambda^3} \; {1 \over 2^{ {3\over 2}} 3^{  {1\over 4}} }~,
\end{equation}
where
\begin{equation}
\Lambda^3  \equiv {\mu^3 \over g^2(\mu)} \; e^{- {2 \pi^2 \over g^2(\mu)}} 
\end{equation}
is the scale parameter of $G_2$ Yang Mills theory, and the explicit 
factors of $g^2$ are normalized at the scale $m_W \sim 1/L$.

 Let us now explain the origin of the various terms in $V_{total}$.
Consider first the supersymmetric case $c_m=0$, when  the potential 
is given by the first two lines in Eq.~(\ref{g2potential}). The terms 
independent of the dual photon $\vec\sigma'$ are due to the three kinds 
of center-stabilizing bions. The terms on the second line are due to 
magnetic bions, which come in two kinds only (because $\alpha_3^* \cdot 
\alpha_2^* =0$). Magnetic bions are responsible for the generation of 
the mass gap of the dual photons and for confinement. It is easy to check
that for $\vec\sigma' = \vec{b}' = 0$ the supersymmetric part of the 
potential vanishes. It is also straightforward to see that this point 
is an extremum and that the masses of $\vec{b}'$ and $\vec\sigma'$ 
excitations are identical, as required by supersymmetry. When the gaugino 
mass is turned on, $c_m \ne 0$, the three kinds of monopole-instantons 
associated with the affine and simple roots generate the terms
in the  potential given on the  last three lines in  Eq.~(\ref{g2potential}). 
These terms can be obtained by a calculation, valid at $m L \ll 1$, 
virtually identical to that of Appendix B of \cite{Poppitz:2012sw}, 
where the $SU(2)$ case was considered.

\begin{figure}[t]
\bc \includegraphics[angle=0, width=0.45\textwidth]{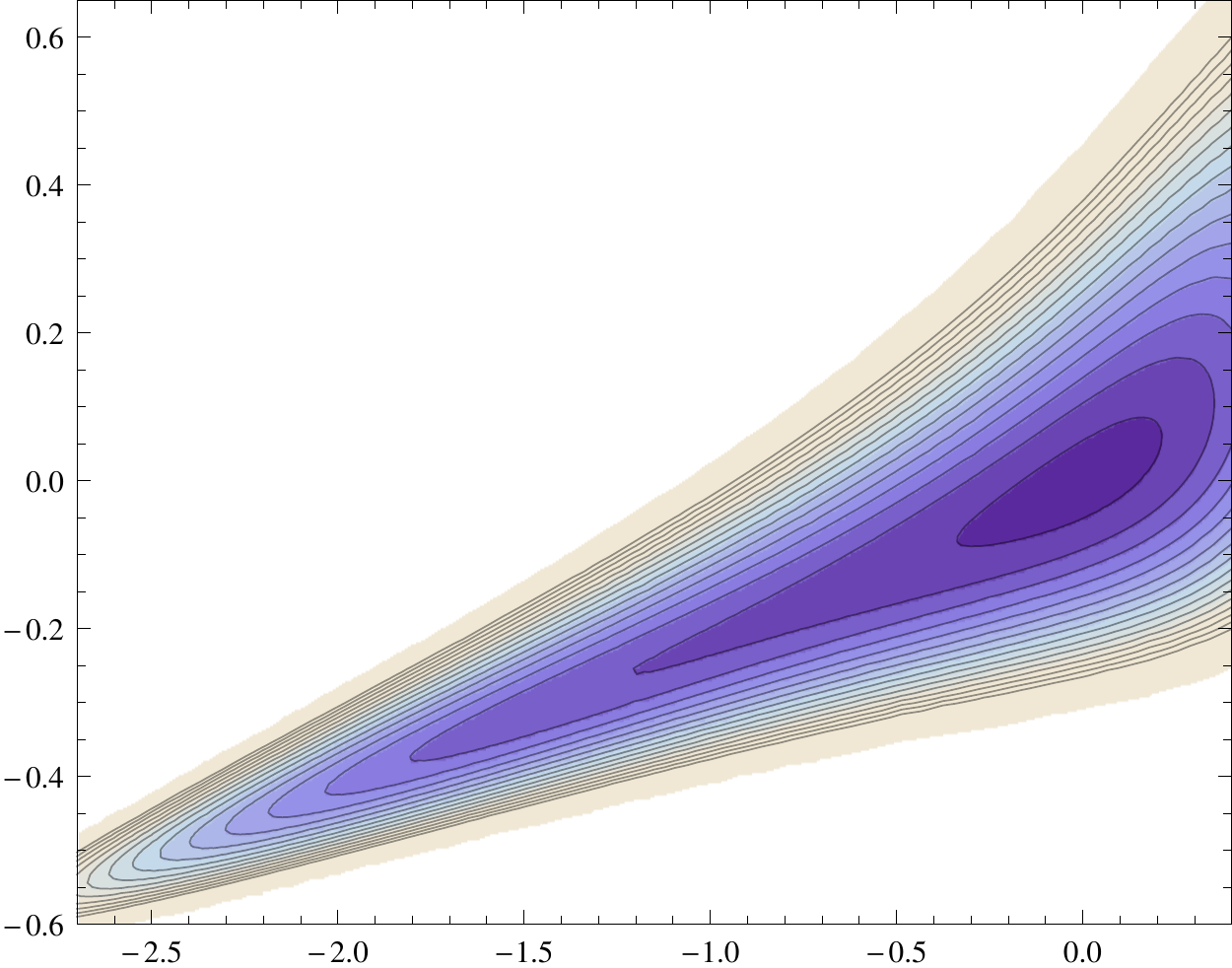}
    \hspace*{0.05\textwidth}
    \includegraphics[angle=0, width=0.45\textwidth]{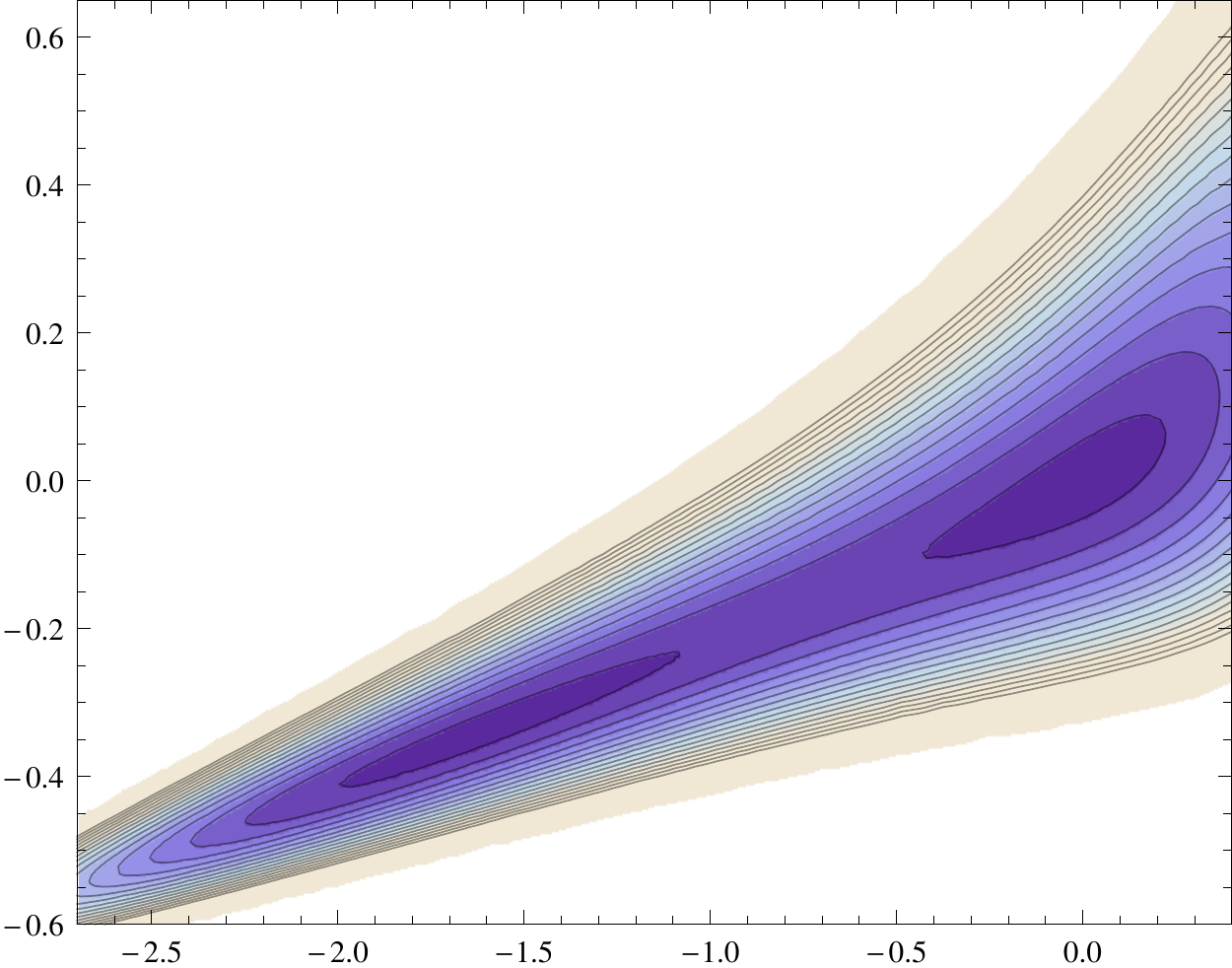}\ec
\caption{\small \label{fig:g2}
Contour plots of the bion- and monopole-instanton-induced potential 
$V_{total}$ for $G_2$ for subcritical and critical values of $c_m$ 
(the gaugino mass), as a function of $b_1'$ and $b_2'$ (at $\theta = 0$). 
Left panel: Contour plot of the potential for $c_m < c_m^{**}< c_m^{\it cr}$ 
($c_m = 1.4$, $c_m^{\it cr} = 1.587$). The minimum at at the origin 
($\langle\vec{b}'\rangle=0, \langle\vec{\sigma}'\rangle=0$) is unique 
and not destabilized by monopole-instantons. The vacuum energy is $-5.6 
V_{bion}^0$.
Right panel: Contour plot of the potential for $c_m$$=$$c_m^{\it cr}=1.587$. 
The minimum at the origin is degenerate with the $\langle\vec{b}'\rangle 
= (-1.648,-0.345)$, $\langle\vec{\sigma}'\rangle = 0$ minimum. The vacuum 
energy is $-6.348 V_{bion}^0$.}
\end{figure}

 Let us now analyze  the minima of the total potential in 
Eq.~(\ref{g2potential}).  In doing so, we shall ignore the ${\cal{O}}(g^2)$ 
terms in the monopole-instanton induced potential. This is consistent in 
the weak-coupling limit. These terms will contribute a shift of the vevs 
of $\vec{b}'$ and $\vec\sigma'$ of order $g^2$, which can further induce 
an order $g^4$ shift in $\langle\Omega\rangle$, which we ignore. As in 
the discussion of the $SU(N_c)$ case, we first study the expansion of 
Eq.~(\ref{g2potential}) to quadratic order around $\langle \vec{b}' 
\rangle = \langle \vec\sigma' \rangle = 0$ and find that a sufficiently 
large $c_m$ (a number of order unity) destabilizes the supersymmetric 
ground state. It is also easy to see that, as in $SU(N_c > 2)$, there 
are cubic terms in the potential for the $\vec{b}'$ fields. Thus, one 
expects a discontinuous transition to a ground state where the field 
$\vec{b}'$ acquires a vev (and perhaps also $\vec\sigma'$). To locate the
transition we study the $\theta = 0$ case in more detail. In this case the  
$\langle \vec{b}' \rangle = \langle \vec\sigma' \rangle = 0$ vacuum with 
$k=0$ has the lowest energy. We find that for $c_m < c_m^{\it cr} \approx 
1.587$ the supersymmetric vacuum is stable. At $c_m = c_m^{\it cr}$, a 
metastable vacuum with $\langle b_1^\prime\rangle = -1.648$, $\langle 
b_2^\prime\rangle  = -0.345$, $\langle \vec\sigma' \rangle = 0$ becomes 
degenerate with the zero-vev ground state. Contour plots of the potential 
are shown in Figure~\ref{fig:g2}.

We find that while there is no symmetry associated with the 
discontinuous transition to a new vacuum at $c_m > c_m^{\it cr}$, the value 
of $\langle\tr\Omega\rangle$ changes discontinuously  from its subcritical 
value (\ref{susypolyakovvev}):
\begin{eqnarray}
\label{g2transition}
c_m < c_m^{\it cr}: && 
     \langle \tr\Omega  \rangle = -0.15 \; {g^2 \over 4 \pi}~, \nonumber \\
c_m = c_m^{{\it cr} +}: && 
     \langle  \tr\Omega \rangle= 3.21 \; {g^2 \over 4 \pi}~.
\end{eqnarray}
To obtain the second number above, we substituted the expectation value 
of $\langle \vec{\phi} \rangle$, Eq.~(\ref{fieldsg2}) with the vevs of 
$\vec{b}'$ as given in the previous paragraph,  into Eq.~(\ref{omegag2}). 
It is interesting to compare Eq.~(\ref{g2transition}) to the results 
obtained in the  lattice simulations of pure Yang Mills $G_2$ theory 
\cite{Cossu:2007dk,Pepe:2006er}. Fig.~4 of Pepe et al.~\cite{Pepe:2006er}
shows histograms of $\langle\tr\Omega\rangle$ in the low and high
temperature phase. The results show that $\langle\tr\Omega \rangle$ 
changes from slightly negative values below $T_c$ to large positive 
values above $T_c$, in agreement with Eq.~(\ref{g2transition}).

\section{Weak vs. strong coupling non-trivial Wilson line holonomy}
\label{sec:weakstrong}

 We would like to conclude with a general discussion of the relation
between semi-classical theories of confinement, discussed in this 
work, and strong coupling confinement, studied on the lattice. For
simplicity we consider $SU(N_c)$ gauge theory.  A question that is 
not well understood is whether the expectation value of the trace of 
the Wilson line  vanishes in the confined phase because

\begin{itemize}
\item[{\it a)}]  the Wilson line is dominated by gauge configurations 
in which its eigenvalues are located at the $N_c$ roots of unity with 
small fluctuations around them. This is the adjoint Higgs regime, 
see Fig.~\ref{fig:phase2}b. 

\item[{\it b)}] fluctuations randomize the eigenvalues over the unit 
circle, and there is no preferred background, as in Fig.~\ref{fig:phase2}c. 
\end{itemize}

This question is a source of confusion especially when one considers 
the phase transition in pure YM theory. There, the transition occurs 
at the strong scale, hence there is no parametric separation of scales 
to justify an effective field theory in the transition regime.\footnote{
An exception is the second order transition of pure $SU(2)$ gauge theory. 
In this case universality arguments imply the existence of a $3d$
effective theory for the Wilson line \cite{Svetitsky:1982gs}. We also 
note that one can always define an effective potential for the Wilson 
line. This potential simply determines the free energy as a function 
of the average Wilson line --- it is not the potential in a local 
effective field theory.} 
This regime is often {\it modelled} by a potential which breaks the center
symmetry in the high temperature deconfined phase and restores it in 
the low temperature confined phase. In the limit of asymptotically high 
$T$  the potential can be justified via a perturbative calculation 
\cite{Gross:1980br}, but at low $T$ the coupling is strong and one 
cannot systematically derive a potential. Ref.~\cite{Dumitru:2012fw} 
discusses this issue and proposes that option {\it a)} is operative 
in the low $T$ confined regime of Yang-Mills theory; note that this point of view is also taken in  \cite{Diakonov:2010qg} and \cite{Shuryak:2012aa}.
  
 First, we emphasize that {\it both} {\it a)} and {\it b)} take place 
in the confined phase of a large class of gauge theories on $\R^3\times 
\S^1_L$, where $\S^1_L$ is a spatial circle. Examples include $\N=1$ SYM 
and QCD(adj), deformed-YM, and deformed-QCD. In all these theories, there 
are two types of confinement operative for different ranges of the 
parameters. We have 
\begin{align}
 \label{scales2}
 &\frac{LN_c\Lambda}{2\pi} \ll 1, \qquad  \text {abelian confinement, 
        {\it a)} is operative,} \nonumber \\
 & \frac{LN_c\Lambda}{2\pi} \gg 1, \qquad  \text{non-abelian confinement, 
        {\it b)} is operative. }
  \end{align}
The small parameter $\frac{LN_c\Lambda}{2\pi} \ll 1$ arises in the abelian 
confinement regime as the separation of scales between the heaviest dual 
photon mass $m_\sigma $ which is degenerate with $m_b$ given in Eq.~(\ref{mb})   
and lightest $W$-boson mass $m_W$:
\begin{align}
\frac{m_\sigma}{m_W} \approx  \left( \frac{\Lambda}{m_W}\right)^3 
    \log  \frac{\Lambda}{m_W} \ll 1~,
\end{align}
implying the first line of Eq.~(\ref{scales2}). This is the regime of 
abelian confinement where an adjoint Higgs mechanism is operative and    
weak coupling non-trivial holonomy shown in Fig.~\ref{fig:phase2}b is 
valid. In this regime, the fluctuations of the eigenvalues are much 
smaller than the typical eigenvalue separation. Once this hierarchy
is lost there is no longer an adjoint Higgs effect: the eigenvalues 
have large fluctuations and are essentially randomized over the unit 
circle. We call this the regime of strong coupling non-trivial holonomy, 
see the discussion in  Ref.~ \cite{Unsal:2008ch}. 
  
  There is some evidence from lattice simulations that the transition 
in pure Yang Mills theory is between a phase with weak coupling trivial 
holonomy as in Fig.~\ref{fig:phase2}a) and a strong coupling non-trivial 
holonomy shown in Fig.~\ref{fig:phase2}c). Fig.~2 of Ref.~\cite{Diakonov:2012dx}
shows the free energy of parameterized Wilson lines $\Omega = {\rm Diag}
(e^{i \pi \nu},e^{-i \pi \nu})$ in pure gauge $SU(2)$ for different temperatures.
At high temperature the free energy agrees with the perturbative one-loop
potential and has a minimum at $\nu=0$. On the other hand,  the free energy in 
the low temperature does not have a pronounced minimum at $\nu =\frac{1}{2}$.
Instead, the free energy is essentially $\nu$ independent, suggesting
that the Wilson line averages to zero because of the large fluctuations 
of $\nu$ in the confined phase. 
   
\subsection{What changes between the semi-classical and strongly coupled 
deconfinement transitions?}

 The phase diagram shown in Fig.~\ref{fig:phasediag} contains both
types of transitions. The transition from the deconfined phase to 
abelian confinement at small $L\Lambda$ continuously evolves to a 
transition to non-abelian confinement at $L\Lambda\sim 1$. Properties 
of the phase transition that can be studied reliably in the abelian 
regime can be extrapolated to the non-abelian regime. One may wonder 
whether one can improve upon this, and find a more direct connection 
between deconfinement transition  in the abelian and non-abelian regime. 

 In the semi-classical limit neutral bions $\cB_{ii}= [\cM_i\overline\cM_i]$ 
provide repulsion between the eigenvalues of the Wilson line. It is not 
hard to see that magnetic bion-anti-bion configurations, 4-defects, such 
as $[\cB_{ij}\overline \cB_{ij}]\equiv [\cB_{ij}\cB_{ji}] $ also generate 
repulsion among eigenvalues, and the same is true for 6-defects $[\cB_{ij}   
\cB_{jk}\cB_{ki}]$. In the semi-classical regime these effects are 
exponentially suppressed relative to bions. Recent works provide
evidence that these molecular configurations are semi-classical 
realizations of IR-renormalons \cite{Argyres:2012vv,Argyres:2012ka}. 
In particular, in the complex Borel-plane, molecular events are 
associated with  singularities located at $t_n^{se.cl.} \sim \frac{2S_I}{N_c}n$, 
$n=(1),2, 3,\ldots$,  where the existence of the $n=1$ term depends on 
details of the particular theory. Note that the location of the singularities 
is smaller by a factor $1/N_c$ relative to the instanton-anti-instanton 
singularity, and that they survive the large-$N_c$ limit discussed in 
Section~\ref{section:large-N-lim}. The location of the infrared renormalon 
poles for the theory on $\R^4$, or for the center-symmetric theory on  
$\R^3 \times \S^1_L$ in the $\frac{LN_c\Lambda}{2\pi} \gg 1$ domain, is 
given by $t_n\sim \frac{2S_I}{\beta_0}n$, $n=(1),2, 3,\ldots$, where 
again, the existence of the $n=1$ term depends on the theory, and 
$\beta_0 =3 N_c$ for SYM and $\beta_0 =\frac{11}{3} N_c$ for YM.  The 
crucial point is that both types of singularities appear with a factor 
of $\frac{1}{N_c}$ up to order one pre-factors. The order one pre-factors 
is what seems to be changing as one moves on from the abelian to 
non-abelian confinement regime, and by adiabatic continuity,  
the IR-renormalons in the strongly coupled phase transition replace
the role of neutral bions in the semi-classical regime.  We believe 
that this direction is worthy of further investigation. 

\section{Prospects}

 In this work, we provide a microscopic mechanism for the deconfinement 
phase transition in a weakly coupled setting which is continuously 
connected to the phase transition in pure gauge Yang Mills theory 
with any gauge group. We find that the mechanism for the phase transition 
is universal, independent of the symmetries of the gauge group. Neutral 
bions generate repulsion among eigenvalues of Wilson line, and this
effect is counter-acted by the perturbative one-loop potential and 
monopole-instantons which generate an attraction. In the weak coupling
regime the calculation is based on the semi-classical expansion, and
the results can be checked (except in the case $\theta\neq 0$) using 
lattice simulations. It may be possible to extend some of the calculations
beyond the semi-classical regime using resurgence theory and transseries, 
see \cite{Argyres:2012ka}.  

 The exact counterpart of our analysis can be carried out to describe 
the deconfinement phase transition or rapid crossover in two-dimensional 
non-linear sigma models, for example the $O(N)$ and $CP(N-1)$ models, or
the chiral principal model. Starting with a supersymmetric version of 
these theories on $\R \times \S^1_L$,  and turning on a mass term for 
the fermions, a  semi-classical transition takes place. We claim that 
the neutral bions (correlated kink-antikink instanton  events in that 
context) are the universal topological configurations leading to the 
confined regime of these theories. 

 There are several possible lines of investigation that can be 
pursued in the future. One is the goal to extend our calculations
into the strongly coupled regime along the lines discussed in the 
previous section, by studying the singularity 
structure in the Borel plane both in the confined and deconfined phases. 
Another direction of study is more detailed 
investigations of confinement in the semi-classical regime, for
example by constructing effective theories for abelian strings,
or for the domain walls that separate the different $k$-vacua. 
In addition to that we are of course interested in experimental 
manifestations of the confinement mechanism. A recent set of ideas
that promises to link deconfinement with the role of topological
objects is the chiral magnetic effect \cite{Fukushima:2008xe}.
This effect leads, in the deconfined phase,  to electric charge separation in a large magnetic 
field,  in the presence of topological 
configurations that can generate net chirality. The effect is 
suppressed by the small density of 4d instantons in the high 
temperature phase, $\sim \exp[-8\pi^2/g^2]$. The density of 
monopole-instantons scales as $\exp[-8\pi^2/(g^2N_c)]$, which may 
result in an enhancement of the chiral magnetic effect. 
 
\acknowledgments

 This work was supported in part by the US Department of Energy grants 
DE-FG02-03ER41260 (T.S.), DE-FG02-12ER41806 (M.\"U.), and the National 
Science and Engineering Research Council of Canada (E.P.). E.P. thanks the Tata Institute for Fundamental Research, Mumbai, for hospitality during the completion of this paper.


\begin{thebibliography}{99}

\bibitem{Poppitz:2012sw} 
E.~Poppitz, T.~Sch\"afer and M.~\" Unsal,
``Continuity, Deconfinement, and (Super) Yang-Mills Theory,''
JHEP {\bf 1210}, 115 (2012)
[arXiv:1205.0290 [hep-th]].

\bibitem{Poppitz:2011wy} 
E.~Poppitz and M.~\"Unsal,
``Seiberg-Witten and Polyakov-like magnetic bion confinements are 
continuously connected,''
JHEP {\bf 1107}, 082 (2011)
[arXiv:1105.3969 [hep-th]].
  
\bibitem{Gross:1980br}
D.~J.~Gross, R.~D.~Pisarski and L.~G.~Yaffe,
``QCD and instantons at finite temperature,''
Rev.\ Mod.\ Phys.\  {\bf 53}, 43 (1981).

\bibitem{Seiberg:1996nz} 
N.~Seiberg and E.~Witten,
``Gauge dynamics and compactification to three-dimensions,''
Proceedings of the conference on mathematical beauty of physics,
Saclay, France (1996), page 333-366
[hep-th/9607163].
  
\bibitem{Aharony:1997bx} 
O.~Aharony, A.~Hanany, K.~A.~Intriligator, N.~Seiberg and M.~J.~Strassler,
``Aspects of N=2 supersymmetric gauge theories in three-dimensions,''
Nucl.\ Phys.\ B {\bf 499}, 67 (1997)
[hep-th/9703110].
  
\bibitem{Davies:2000nw} 
N.~M.~Davies, T.~J.~Hollowood and V.~V.~Khoze,
``Monopoles, affine algebras and the gluino condensate,''
J.\ Math.\ Phys.\  {\bf 44}, 3640 (2003)
[hep-th/0006011].

\bibitem{Unsal:2010qh} 
M.~\"Unsal and L.~G.~Yaffe,
``Large-N volume independence in conformal and confining gauge theories,''
JHEP {\bf 1008}, 030 (2010)
[arXiv:1006.2101 [hep-th]].

\bibitem{Lee:1997vp}
K.~-M.~Lee, P.~Yi,
``Monopoles and instantons on partially compactified D-branes,''
Phys.\ Rev.\  {\bf D56}, 3711-3717 (1997).
[hep-th/9702107].

\bibitem{Kraan:1998sn}
T.~C.~Kraan and P.~van Baal,
``Monopole constituents inside $SU(n)$ calorons,''
Phys.\ Lett.\  B {\bf 435}, 389 (1998)
[arXiv:hep-th/9806034].

\bibitem{Nye:2000eg}
T.~M.~W.~Nye and M.~A.~Singer,
``An ${\cal{L}}$$^2$-index theorem for Dirac operators 
on $\R^3 \times \S^1$,'' 
J.\ Funct.\ Anal.\ {\bf 177}, 203 (2000)
[arXiv:math/0009144].
  
\bibitem{Poppitz:2008hr}
E.~Poppitz and M.~\" Unsal,
``Index theorem for topological excitations on  $\R^3 \times \S^1$ and 
Chern-Simons theory,''
JHEP {\bf 0903}, 027 (2009)
[arXiv:0812.2085 [hep-th]].
   
\bibitem{Unsal:2007jx}
M.~\" Unsal,
``Magnetic bion condensation: A new mechanism of confinement and mass gap in
four dimensions,''
Phys.\ Rev.\  D {\bf 80} (2009) 065001
[arXiv:0709.3269 [hep-th]].

\bibitem{Anber:2011de}
M.~M.~Anber, E.~Poppitz,
``Microscopic structure of magnetic bions,''
JHEP {\bf 1106}, 136 (2011).
[arXiv:1105.0940 [hep-th]].

\bibitem{Argyres:2012ka} 
P.~C.~Argyres,  and M.~\"Unsal,
``The semi-classical expansion and resurgence in gauge theories: 
new perturbative, instanton, bion, and renormalon effects,''
JHEP {\bf 1208}, 063 (2012)
[arXiv:1206.1890 [hep-th]].


\bibitem{Lucini:2005vg} 
  B.~Lucini, M.~Teper and U.~Wenger,
  ``Properties of the deconfining phase transition in SU(N) gauge theories,''
  JHEP {\bf 0502}, 033 (2005)
  [hep-lat/0502003].

\bibitem{Panero:2009tv} 
  M.~Panero,
 ``Thermodynamics of the QCD plasma and the large-N limit,''
  Phys.\ Rev.\ Lett.\  {\bf 103}, 232001 (2009)
  [arXiv:0907.3719 [hep-lat]].

\bibitem{Mykkanen:2012ri} 
  A.~Mykkanen, M.~Panero and K.~Rummukainen,
 ``Casimir scaling and renormalization of Polyakov loops in large-N gauge theories,''
  JHEP {\bf 1205}, 069 (2012)
  [arXiv:1202.2762 [hep-lat]].

\bibitem{Lucini:2012gg} 
B.~Lucini and M.~Panero,
``SU(N) gauge theories at large N,''
arXiv:1210.4997 [hep-th].
  
\bibitem{Beringer:1900zz} 
J.~Beringer {\it et al.}  [Particle Data Group Collaboration],
``Review of Particle Physics (RPP),''
Phys.\ Rev.\ D {\bf 86}, 010001 (2012).

\bibitem{Witten:1998uka} 
E.~Witten,
``Theta dependence in the large N limit of four-dimensional gauge theories,''
Phys.\ Rev.\ Lett.\  {\bf 81}, 2862 (1998)
[hep-th/9807109].

\bibitem{Unsal:2008ch} 
M.~\" Unsal and L.~G.~Yaffe,
``Center-stabilized Yang-Mills theory: Confinement and large N volume 
independence,''
Phys.\ Rev.\ D {\bf 78}, 065035 (2008)
[arXiv:0803.0344 [hep-th]].

\bibitem{Bringoltz:2005xx} 
B.~Bringoltz and M.~Teper,
``In search of a Hagedorn transition in SU(N) lattice gauge theories at 
large-N,''
Phys.\ Rev.\ D {\bf 73}, 014517 (2006)
[hep-lat/0508021].

\bibitem{Cabibbo:1975ig} 
N.~Cabibbo and G.~Parisi,
``Exponential Hadronic Spectrum and Quark Liberation,''
Phys.\ Lett.\ B {\bf 59}, 67 (1975).

\bibitem{Thorn:1980iv} 
C.~B.~Thorn,
``Infinite $N_c$ QCD at finite temperature: Is there an ultimate temperature?,''
Phys.\ Lett.\ B {\bf 99}, 458 (1981).

\bibitem{Cohen:2006qd} 
T.~D.~Cohen,
``QCD strings and the thermodynamics of the metastable phase of QCD at 
large N(c),''
Phys.\ Lett.\ B {\bf 637}, 81 (2006)
[hep-th/0602037].

\bibitem{Aharony:2003sx} 
O.~Aharony, J.~Marsano, S.~Minwalla, K.~Papadodimas and M.~Van Raamsdonk,
``The Hagedorn - deconfinement phase transition in weakly coupled large N 
gauge theories,''
Adv.\ Theor.\ Math.\ Phys.\  {\bf 8}, 603 (2004)
[hep-th/0310285].

\bibitem{Aharony:2005bq} 
O.~Aharony, J.~Marsano, S.~Minwalla, K.~Papadodimas and M.~Van Raamsdonk,
``A First order deconfinement transition in large N Yang-Mills theory on 
a small S**3,''
Phys.\ Rev.\ D {\bf 71}, 125018 (2005)
[hep-th/0502149].
  
\bibitem{Myers:2007vc} 
J.~C.~Myers and M.~C.~Ogilvie,
``New phases of SU(3) and SU(4) at finite temperature,''
Phys.\ Rev.\ D {\bf 77}, 125030 (2008)
[arXiv:0707.1869 [hep-lat]].
  
\bibitem{Ogilvie:2012is} 
M.~C.~Ogilvie,
``Phases of Gauge Theories,''
J.\ Phys.\ A {\bf 45}, 483001 (2012)
[arXiv:1211.2843 [hep-th]].
  
\bibitem{D'Elia:2012vv} 
M.~D'Elia and F.~Negro,
``$\theta$ dependence of the deconfinement temperature in Yang-Mills 
theories,''
Phys.\ Rev.\ Lett.\  {\bf 109}, 072001 (2012)
[arXiv:1205.0538 [hep-lat]].

\bibitem{Vicari:2008jw} 
E.~Vicari and H.~Panagopoulos,
``Theta dependence of SU(N) gauge theories in the presence of a topological 
term,''
Phys.\ Rept.\  {\bf 470}, 93 (2009)
[arXiv:0803.1593 [hep-th]].
  
\bibitem{Thomas:2011ee} 
  E.~Thomas and A.~R.~Zhitnitsky,
``Topological Susceptibility and Contact Term in QCD. A Toy Model,"
  Phys.\ Rev.\ D {\bf 85}, 044039 (2012)
  [arXiv:1109.2608 [hep-th]].

\bibitem{Unsal:2012zj} 
  M.~\"Unsal,
  ``Theta dependence, sign problems and topological interference,''
  Phys.\ Rev.\ D {\bf 86}, 105012 (2012)
  [arXiv:1201.6426 [hep-th]].
  
\bibitem{Anber:2013sga} 
  M.~M.~Anber,
``Theta dependence of the deconfining phase transition in pure $SU(N_c)$ Yang-Mills theories,"
  arXiv:1302.2641 [hep-th].
  
\bibitem{Holland:2003jy} 
K.~Holland, P.~Minkowski, M.~Pepe and U.~J.~Wiese,
``Exceptional confinement in G(2) gauge theory,''
Nucl.\ Phys.\ B {\bf 668}, 207 (2003)
[hep-lat/0302023].

\bibitem{Pepe:2006er} 
M.~Pepe and U.~-J.~Wiese,
``Exceptional Deconfinement in G(2) Gauge Theory,''
Nucl.\ Phys.\ B {\bf 768}, 21 (2007)
[hep-lat/0610076].

\bibitem{Greensite:2006sm} 
J.~Greensite, K.~Langfeld, S.~Olejnik, H.~Reinhardt and T.~Tok,
``Color Screening, Casimir Scaling, and Domain Structure in G(2) and 
SU(N) Gauge Theories,''
Phys.\ Rev.\ D {\bf 75}, 034501 (2007)
[hep-lat/0609050].

\bibitem{Cossu:2007dk} 
G.~Cossu, M.~D'Elia, A.~Di Giacomo, B.~Lucini and C.~Pica,
``G(2) gauge theory at finite temperature,''
JHEP {\bf 0710}, 100 (2007)
[arXiv:0709.0669 [hep-lat]].

\bibitem{Diakonov:2010qg} 
D.~Diakonov and V.~Petrov,
``Confinement and deconfinement for any gauge group from dyons viewpoint,''
AIP Conf.\ Proc.\  {\bf 1343}, 69 (2011)
[arXiv:1011.5636 [hep-th]].

\bibitem{Ilgenfritz:2012wg} 
E.~-M.~Ilgenfritz and A.~Maas,
``Topological aspects of G2 Yang-Mills theory,''
arXiv:1210.5963 [hep-lat].
  
\bibitem{Dumitru:2012fw} 
A.~Dumitru, Y.~Guo, Y.~Hidaka, C.~P.~K.~Altes and R.~D.~Pisarski,
``Effective Matrix Model for Deconfinement in Pure Gauge Theories,''
Phys.\ Rev.\ D {\bf 86}, 105017 (2012)
[arXiv:1205.0137 [hep-ph]].

\bibitem{Shuryak:2012aa} 
  E.~Shuryak and T.~Sulejmanpasic,
``The Chiral Symmetry Breaking/Restoration in Dyonic Vacuum,"
  Phys.\ Rev.\ D {\bf 86}, 036001 (2012)
  [arXiv:1201.5624 [hep-ph]].
  

\bibitem{Svetitsky:1982gs} 
B.~Svetitsky and L.~G.~Yaffe,
``Critical Behavior at Finite Temperature Confinement Transitions,''
Nucl.\ Phys.\ B {\bf 210}, 423 (1982).

\bibitem{Diakonov:2012dx} 
D.~Diakonov, C.~Gattringer and H.~-P.~Schadler,
``Free energy for parameterized Polyakov loops in SU(2) and SU(3) 
lattice gauge theory,''
JHEP {\bf 1208}, 128 (2012)
[arXiv:1205.4768 [hep-lat]].

\bibitem{Argyres:2012vv} 
P.~Argyres and M.~\"Unsal,
``A semiclassical realization of infrared renormalons,''
Phys.\ Rev.\ Lett.\  {\bf 109}, 121601 (2012)
[arXiv:1204.1661 [hep-th]].


\bibitem{Fukushima:2008xe} 
K.~Fukushima, D.~E.~Kharzeev and H.~J.~Warringa,
``The Chiral Magnetic Effect,''
Phys.\ Rev.\ D {\bf 78}, 074033 (2008)
[arXiv:0808.3382 [hep-ph]].
  

  
\end{thebibliography}
\end{document}